\documentclass[useAMS,usenatbib,article]{mn2e}
\usepackage[dvips]{graphicx}
\usepackage{longtable}
\title[Infrared Dark Cloud Cores in the SCUBA Legacy Catalogue]{Infrared Dark Cloud Cores in the SCUBA Legacy Catalogue}
\author[H. Parsons, M. A. Thompson and A. Chrysostomou]{H. Parsons$^{1}$\thanks{E-mail: 
H.A.L.Parsons@Herts.ac.uk;}, M. A. Thompson$^{1}$\thanks{E-mail: 
M.A.Thompson@Herts.ac.uk;} and A. Chrysostomou$^{1,2}$\\
$^{1}$Centre for Astrophysics Research, Science \& Technology Research Institute, \\
University of Hertfordshire, College Lane, Hatfield, AL10 9AB, UK \\
$^{2}$Joint Astronomy Centre, 660 North A'ohoku Place, University Park, Hilo, Hawaii 96720, U.S.A. }

\begin{document}

\date{Accepted 2009 July 7.  Received 2009 July 7; in original form 2009 March 13}
\pagerange{\pageref{firstpage}--\pageref{lastpage}} \pubyear{2009}
\maketitle

\label{firstpage}

\begin{abstract}
We present an investigation of candidate Infrared Dark Cloud cores as identified by \citet{2006ApJ...639..227S} located within the SCUBA Legacy Catalogue.  After applying a uniform noise cut to the Catalogue data we identify 154 Infrared Dark Cloud cores that were detected at 850\,$\umu$m and 51 cores that were not. We derive column densities for each core from their 8\,$\umu$m extinction and find that the IRDCs detected at 850\,$\umu$m have higher column densities (a mean of $1.7\times10^{22}$\,cm$^{-2}$) compared to those cores not detected at 850\,$\umu$m (a mean of $1.0\times10^{22}$\,cm$^{-2}$). Combined with sensitivity estimates, we suggest that the cores not detected at 850\,$\umu$m are low mass, low column density and low temperature cores that are below the sensitivity limit of SCUBA at 850\,$\umu$m. For a subsample of the cores detected at 850\,$\umu$m those contained within the MIPSGAL area) we find that two thirds are associated with 24\,$\umu$m sources. Cores not associated with 24\,$\umu$m emission are either ``starless'' IRDC cores that perhaps have yet to form stars, or contain low mass YSOs below the MIPSGAL detection limit. We see that those ``starless'' IRDC cores and the IRDC cores associated with 24\,$\umu$m emission are drawn from the same column density population and are of similar mass. If we then assume the cores without 24\,$\umu$m embedded sources are at an earlier evolutionary stage to cores with embedded objects we derive a statistical lifetime for the quiescent phase of a few 10$^{3}$--10$^{4}$\,years. Finally, we make conservative predictions for the number of observed IRDCs that will be observed by the Apex Telescope Galactic Plane Survey (ATLASGAL), the Herschel Infrared Galactic Plane Survey (Hi-GAL), the JCMT Galactic Plane Survey (JPS) and the SCUBA-2 ``All Sky'' Survey (SASSy).
\end{abstract}

\begin{keywords}
stars: formation -- dust --infrared: ISM -- submillimetre
\end{keywords}


\section{Introduction}

Infrared Dark Clouds (IRDCs) were first observed in the mid-1990s by the Infrared Space Observatory, ISO, \citep{1996A&A...315L.165P} and the Midcourse Space eXperiment, MSX, \citep{1998ApJ...494L.199E} as silhouettes against the bright mid-infrared Galactic background. Initially, \cite{1998ApJ...494L.199E} identified $\sim$2000 clouds by eye from the MSX Galactic Plane Survey images. A systematic study of the MSX data using an automated identification process by \cite{2006ApJ...639..227S} identified 10,931 candidate IRDCs within which a total of 12,774 compact IRDC core candidates were detected. By definition each IRDC contains at least one core.

An unsurprising consequence of this detection method is the fact that the Galactic distribution of IRDCs follows the mid-infrared background of the Galaxy. IRDCs are predominantly found in the first and fourth Galactic quadrants and near to the Galactic mid-plane \citep{2008ApJ...680..349J}, precisely where the mid-infrared background is greatest. \citet{2006ApJ...653.1325S} used data from the Galactic Ring Survey (GRS) to obtain distance estimates for IRDCs found within the first quadrant of the Galaxy, using a morphological match with $^{13}$CO emission. This matching process identified distances to 313 candidate IRDCs. Further distance estimates have been obtained by \citet{2008ApJ...680..349J} for 316 IRDC candidates contained within the fourth quadrant and were obtained using single pointings of CS (J\,=\,2--1). The distances obtained from these large scale investigations show a peak in the radial galactocentric distribution of IRDCs corresponding to the location of the Scutum-Centaurus arm i.e. peak of R\,=\,5kpc in the first quadrant and R\,=\,6kpc in the fourth \citep{2008ApJ...680..349J}.

In the decade since their discovery, our understanding of the physical properties of IRDCs has increased and it is now known that these objects are cold ($<$25\,K) dense (10$^{5}$\,cm$^{-3}$) regions, on scales of 1--10\,pc, with masses ranging between 10$^{2}$-10$^{5}$\,M$_{\odot}$ \citep{2006ApJ...641..389R}. Current theory suggests that cold dense starless cores found within IRDCs are the precursors to hot molecular cores (\citealt{2008ASPC..387...44J} and references therein), indeed \citet{2007ApJ...662.1082R} report on a hot molecular core found within an IRDC. Other tracers of massive star formation such as HII regions and Class II methanol (CH$_{3}$OH) masers (\citealt{1998ApJ...508..721C}; \citealt{2006A&A...447..929P}; \cite{2009ApJS..181..360C}) have been found in association with a number of IRDCs. The association with high mass star formation is not exclusive: a number of individual studies find only low to intermediate mass young stellar objects embedded within IRDCs \citep{2008arXiv0807.3628V}.

To date, with the exception of distance estimates, only small samples of the IRDCs originally published by \citet{2006ApJ...639..227S} have been investigated, with an observational bias towards the darkest high contrast clouds (\citealt{2006ApJ...641..389R}; \citealt{2008ApJ...686..384D}).
The trends of global properties across a large sample of IRDCs have yet to be investigated, in particular the proportion of IRDCs that are associated with active star formation as opposed to IRDCs that are quiescent or starless.

This paper aims to address this issue by studying candidate IRDC cores originally identified by \citet{2006ApJ...639..227S} that are contained within the recently published SCUBA Legacy Catalogue by \citet{2008ApJS..175..277D}. In Section \ref{Section:CrossMatch} we describe the cross matching method used on the two catalogues, obtain column density and mass estimates (or upper limits where applicable) and identify 24\,$\umu$m embedded sources associated with the the cores identified. The results of the cross matching procedure are presented in Section \ref{Section:Results}. We discuss our findings in Section \ref{Section:Discussion} with a mention of the impact on two of the forthcoming James Clerk Maxwell Telescope\footnote{The James Clerk Maxwell Telescope is operated by the Joint Astronomy Centre on behalf of the Scientific and Technology Facilities Council of the UK, the Netherlands Association for Scientific Research, and the National Research Council of Canada.} (JCMT) Legacy surveys. Finally we make some concluding remarks in Section \ref{Section:Conclusions}.


\section{Cross-matching IRDCs in MSX, SCUBA and MIPSGAL}
\label{Section:CrossMatch}
\subsection{Archival data}

In \citeyear{2006ApJ...639..227S}, \citeauthor{2006ApJ...639..227S} produced a catalogue of 10,931 candidate IRDCs using data from the MSX satellite that covered the entire Galactic plane from l\,=\,0\,--\,360$^{\circ}$ and $|$b$|$\,$\le$\,5$^{\circ}$. Candidate IRDCs were identified by modelling the Galactic background diffuse emission at 8\,$\umu$m, subtracting the 8\,$\umu$m MSX images from this model and then dividing by the background model to produce what is known as a ``contrast image''. Regions of high extinction in the raw images appeared as positive objects with contrast values between 0 and 1 (1 for highly extincted objects) in the contrast images. IRDCs were then identified by looking for extended contrast sources, those with 36 or more continuous pixels with a contrast greater than 2$\sigma$. Cores within the clouds were identified by decomposing the clouds using two-dimensional elliptical Gausian fits \citep{2006ApJ...639..227S}. Although discovered by their mid-infrared absorption, it is at sub millimetre and far-infrared wavelengths that these objects have their peak emission.

\citet{2008ApJS..175..277D} present a comprehensive re-reduction of the entire 8 year sub millimetre continuum data set observed by SCUBA (Sub millimetre Common User Bolometer Array) on the JCMT in Hawaii. This data set is known as the SCUBA Legacy Catalogue and covers a total area of just over 29 square degrees at 850\,$\umu$m. A consequence of the varying weather conditions and method by which the data was collected, over the entire lifetime of SCUBA, is that the data within the SCUBA Legacy Catalogue is both non-uniform in noise and in its quality of opacity corrections. \citet{2008ApJS..175..277D} divided the SCUBA Legacy Catalogue into a Fundamental and an Extended Dataset. The former uses data for which there is well known atmospheric opacity calibration data (from both skydips and the CSO radiometer, \citealt{2008ApJS..175..277D}) and the latter contains all observations regardless of the data quality. Coverage by the Extended data set is greater in area than the Fundamental by 9.7 square degrees. Discrete objects were identified within the SCUBA Legacy Catalogue (from both the Fundamental and Extended Dataset independently) using Clumpfind\footnote{The Clumpfind algorithm used to identify objects was adapted from \cite{1994ApJ...428..693W} \citep{2008ApJS..175..277D}.} \citep{2008ApJS..175..277D}. This process provides information on the properties of each object such as flux (at the peak and integrated over its area) and apparent size. 

MIPSGAL\footnote{MIPS (Multiband Imaging Photometer for Spitzer) Galactic Plane Survey. Data available from http://irsa.ipac.caltech.edu/data/SPITZER/MIPSGAL/images/} is a survey of the Galactic Plane from 10\,$<$\,l\,$<$\,65$^{\circ}$ and -10\,$>$\,l\,$>$\,-65$^{\circ}$ $|$b$|$\,$<$\,1$^{\circ}$ at 24 and 70\,$\umu$m, using the MIPS instrument on Spitzer\footnote{The Spitzer Space Telescope is operated by the Jet Propulsion Laboratory, California Institute of Technology under NASA contract 1407.}. As a tracer of warm dust and with good resolution (6" angular resolution as opposed to 20" and 14" for MSX and SCUBA respectively), MIPSGAL 24\,$\umu$m data are ideal for investigating warm embedded objects, such as Young Stellar Objects. Indeed \cite{2008arXiv0808.2053F} combined MIPSGAL data with IRAC (Infrared Camera on Spitzer), 2MASS (Two Micron All Sky Survey) and SCUBA data to identify Young Stellar Objects (YSOs) within IRDC MSXDC G048.65-00.29.

\subsection{Cross identification}
\label{subsection:crossidentification}

In total, 428 MSX IRDC cores from \cite{2006ApJ...639..227S} were located in regions mapped by SCUBA (325 located in the Fundamental region and an additional 103 located in the Extended region). However the SCUBA Legacy Catalogue does not contain photometric measurements of objects located at the edges of maps (which may be subject to large scale background or noise fluctuations). In addition, regions of high noise persist within the catalogue due to the non-uniform way in which the data were taken. In order to make the catalogue more uniform we exclude data with an rms noise greater than 0.1\,Jy\,pixel$^{-1}$, the pixel size for data in the SCUBA Legacy Catalogue is 6''. This 0.1\,Jy\,pixel$^{-1}$ cut excludes the high noise Poisson tail present in the data \citep{2008ApJS..175..277D} and gives the remaining data approximately Gaussian noise statistics, with a mean and sigma of 0.05\,Jy\,pixel$^{-1}$ and 0.024\,Jy\,pixel$^{-1}$ respectively. Taking map edges into account and excluding regions with noise $>$\,0.1\,Jy\,pixel$^{-1}$ leaves us with 251 MSX IRDC cores within the SCUBA Legacy Catalogue mapped region. The 251 MSX candidate IRDC cores were then cross matched to sources identified within the SCUBA Legacy Catalogue (as defined by Clumpfind). IRDC cores were matched against both the Fundamental and the Extended data Catalogues.

For the cross identification process, the locations of the MSX IRDC cores were positionally matched using TOPCAT\footnote{TOPCAT: Tool for OPerations on Catalogues And Tables. See http://www.star.bris.ac.uk/$\sim$mbt/topcat/} to the locations of the Clumpfind SCUBA objects. The irregular morphology of the candidate objects in both the MSX 8\,$\umu$m contrast images and the SCUBA 850\,$\umu$m emission maps, meant that the task of cross matching cores between the two catalogues was non trivial. \citet{2006ApJ...639..227S} identified IRDCs within the MSX 8\,$\umu$m data as contiguous structures in the contrast images that were sufficently extended to be real clouds and not artifacts. The cores within the clouds were identified using two dimensional elliptical Gaussian fits to these contiguous structures. In contrast \citet{2008ApJS..175..277D} created the SCUBA Legacy Catalogue using Clumpfind which identifies irregular objects by following intensity contours. Due to this different approach in identification between the two catalogues the catalogued positions of IRDC cores and SCUBA clumps may differ by a considerable amount, even when the two are clearly morphologically associated with each other. A large positional matching radius was required to identify potential matches followed by further refinement by eye, checking that the individual IRDC cores were morphologically similar to the SCUBA 850\,$\umu$m emission. A matching radius of 1' was chosen as \citet{2006ApJ...639..227S} quotes that typical core diameters lie between 0.75' and 2'. For added confidence, those cores that were initially matched were additionally checked for 850\,$\umu$m emission at the location of the MSX core.

\label{subsection:embedded}
\begin{figure}
\includegraphics[width=0.4\textwidth,trim=0 0 0 30,clip]{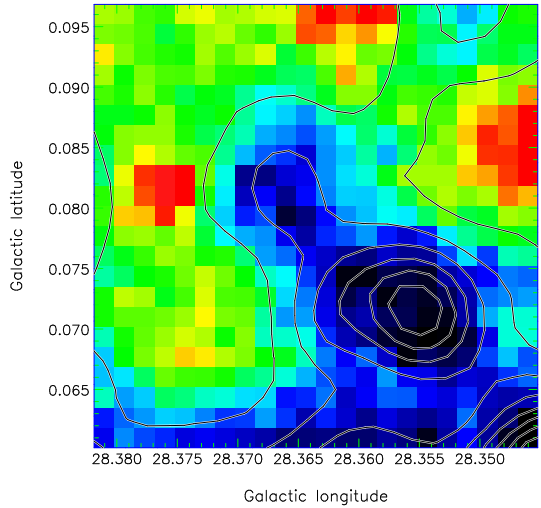}
\caption{Image of IRDC core: MSXDCG028.37+00.07 (d). Image is MSX 8\,$\umu$m with SCUBA contours overlaid.}
 \label{pic:limitation}
\end{figure}

On several occasions multiple matches were made to the same SCUBA sources where Clumpfind had only identified one object and vice versa. In these instances the closest positional Clumpfind match to the candidate MSX identified IRDC core was taken. A consequence of the differing techniques used to identify objects within the MSX and SCUBA catalogues was seen when classifying by eye those IRDC SCUBA detected candidates with and without embedded objects. Fig. \ref{pic:limitation}, shows IRDC core MSXDCG028.37+00.07 (d), Clumpfind identified two distinct objects in the 850\,$\umu$m data but the MSX identification process identified the dark complex as one object.

Of the 251 IRDC cores located within the SCUBA Legacy Catalogue mapped region a total of 46 core matches were manually excluded from the sample. In some cases, this was because the MSX identified IRDC cores were located on positions of extended 850\,$\umu$m emission which could not be morphologically matched to the compact candidate IRDC cores. Due to the large upper limit used for the matching radius a number of matches were also found to be inappropriate. In other cases, objects were found within 1' but not coincident with 850\,$\umu$m emission. This may have occurred due to poor background modelling of the mid-infrared emission. The subtraction of a smoothed background model can potentially result in the creation of artifacts with high contrast values. Cores adjacent to bright extended 8\,$\umu$m infrared emission, were also amongst those excluded. These objects were removed due to concerns over the process of creating a contrast image in a complex enviroment.

Finally of the 205 remaining IRDC cores, a total of 154 IRDC cores were matched to SCUBA objects from the SCUBA Legacy Catalogue of \citet{2008ApJS..175..277D}. The other remaining 51 IRDC cores were identified as having no corresponding 850\,$\umu$m emission. These MSX identified candidate cores could be due to column densities and dust temperatures below the detection limit of SCUBA. Alternatively they could be a result of uncertainties in the MSX IRDC candidate identification process, we explore these possibilities further in section \ref{Section:Discussion1}.

\subsection{Column densities and masses of the cores}
\label{subsection:coldenmass}

\begin{figure}
 \includegraphics[width=0.4\textwidth,angle=-90]{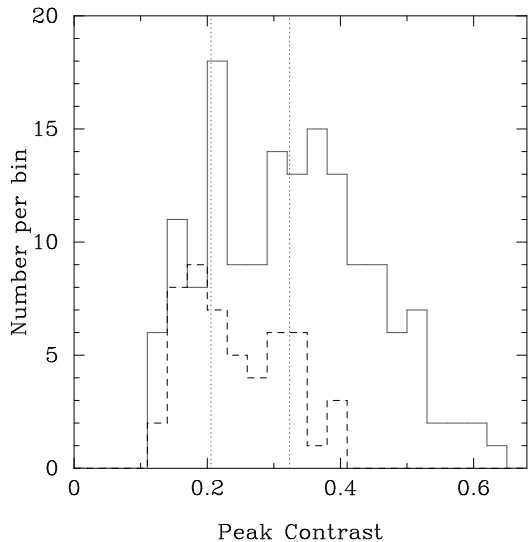}
 \caption{Histogram of Peak Contrast values for cores detected at 850\,$\umu$m (solid line) and cores not detected at 850\,$\umu$m (dashed line). Vertical lines mark the position of the peak contrast values that correspond to $1.7\times 10^{22}$\,cm$^{-2}$ and $1.0\times 10^{22}$\,cm$^{-2}$ (median column densities for cores detected and not detected at 850\,$\umu$m respectively).
Bin size used is 0.03.}
 \label{Hist:ColDen}
\end{figure}

We derive peak column densities for all IRDC cores within our sample, whether detected at 850\,$\umu$m or not, by applying the following extinction law to the MSX 8\,$\umu$m data:
\begin{equation}
I_{i} = I_{b} e^{-\tau_{\lambda}}
\label{equation1}
\end{equation}
where $I_{i}$ is the image intensity, $I_{b}$ is the background model intensity and $\tau_{\lambda}$ is the dust opacity which equals the cross sectional area, $\sigma_{\lambda}$, multiplied by the column density, $N(H_{2})$, i.e. $\tau_{\lambda}=\sigma_{\lambda}.N(H_{2})$. The peak contrast value $C$ is defined by \cite{2006ApJ...639..227S} as: \begin{equation}
C = \frac{I_{b} - I_{i}}{I_{b}} =  1 - e^{-\tau_{\lambda}}
\label{equation:C}
\end{equation} It is then possible, by substituting $\tau_{\lambda}$ from equation \ref{equation1} into equation \ref{equation:C}, to derive:
\begin{equation}
N(H_{2}) = \frac{-ln (1 - C)}{\sigma_{\lambda}}
\label{equation:N8(H2)}
\end{equation} 
We assume a value of $\sigma_{\lambda} = 2.3 \times 10^{-23}$\,cm$^{2}$ for the cross sectional area of the obscuring dust particles at 8.8\,$\umu$m \citep{2006ApJS..166..567R}. Column densities for the IRDCs derived by this method are contained in Tables \ref{table:classA} and \ref{table:classC}. Fig. \ref{Hist:ColDen} shows the distribution of those cores detected and not detected at 850\,$\umu$m with peak contrast. We see that the median column density for the SCUBA detected candidates and the SCUBA non-detected candidates is $1.7\times 10^{22}$\,cm$^{-2}$ and $1.0\times 10^{22}$\,cm$^{-2}$ respectively.

For comparison with the values determined from the 8\,$\umu$m extinction we also calculated the peak column densities for cores detected at 850\,$\umu$m, using the SCUBA 850\,$\umu$m data to derive the mass (as defined by \citeauthor{1983QJRAS..24..267H} \citeyear{1983QJRAS..24..267H}) and assuming spherical geometry:

\begin{equation}
N(H_{2}) = \frac{F_{\nu} C_{\nu}}{B_{\nu}(T) \pi (tan(B_{850}))^{2} 2 m_{H}}
\label{equation:N850(H2)}
\end{equation}
where the mass coefficient $C_\nu$ is given by
\begin{equation}
C_{\nu} = \frac{M_{g}}{M_{d} \kappa_{\nu}}
\end{equation} $F_{\nu}$ is the observed peak flux;  $B_{\nu}(T)$ is the Planck function evaluated for dust temperature, $T$; $B_{850}$ is the radius of the beam at 850\,$\umu$m, which has a FWHM of 19'' due to convolution during the data reduction \citep{2008ApJS..175..277D} and $m_{H}$ is the mass of a hydrogen atom. The value of C$_{\nu}=$ 50\,g\,cm$^{-2}$ at 850\,$\umu$m is taken from \cite{2001ApJ...552..601K}, where $M_{g}$ is the gas mass and $M_{d}$ is the dust mass and $\kappa_{\nu}$ is the dust opacity, assuming a gas to dust ratio of 100 and an opacity gradient $\beta$ of 2.

When evaluating the Planck function, a temperature of 15\,K was assumed for all the cores as this is the midpoint of the observed range (8--25\,K) in IRDC temperatures observed by \citet{1998ApJ...508..721C}, \citet{2002A&A...382..624T} and \citet{2006A&A...450..569P}. Decreasing or increasing the temperature to 8 or 25\,K would increase or decrease these column density estimates by a factor of 3.5 and 2.2 respectively.

We compared the column densities derived by each method. In general the column density derived from the 8\,$\umu$m extinction agrees with that derived from the 850\,$\umu$m emission to within an order of magnitude. There is considerable scatter but the overall trend is that the 8\,$\umu$m column density underestimates the 850\,$\umu$m column density by roughly a factor of 2. This suggests that the average temperature for the IRDC cores may be closer to 10\,K than our assumption of 15\,K. However due to the large uncertainties in mass coefficients, the 8\,$\umu$m extinction law and contamination from foreground emission we do not expect close agreement between these two methods.

Masses were determined for the cores detected at 850\,$\umu$m using the method of \citet{1983QJRAS..24..267H}, i.e.
\begin{equation}
M = \frac{d^{2} F_{\nu} C_{\nu}}{B_{\nu}(T)}
\label{mass}
\end{equation} A dust temperature of 15\,K was again assumed for all the cores. As before, decreasing or increasing the dust temperature to 8 or 25\,K would increase or decrease the masses derived by a factor of 3.5 and 2.2 respectively. $d$ is the distance to each core. Kinematic distances exist for 33 of our cores detected at 850\,$\umu$m from \cite{2006ApJ...639..227S}, who derived distances by matching up the morphologies of candidate IRDC cores with CO morphologies from the GRS (Galactic Ring Survey). Our mass estimates for these 33 SCUBA detected cores, along with the distance estimates from \cite{2006ApJ...639..227S} can be found in Table \ref{table:mass}.

\subsection{Embedded 24\,$\umu$m objects in the cores}
\begin{figure*}
\includegraphics[width=0.4\textwidth,trim=0 0 0 30,clip]{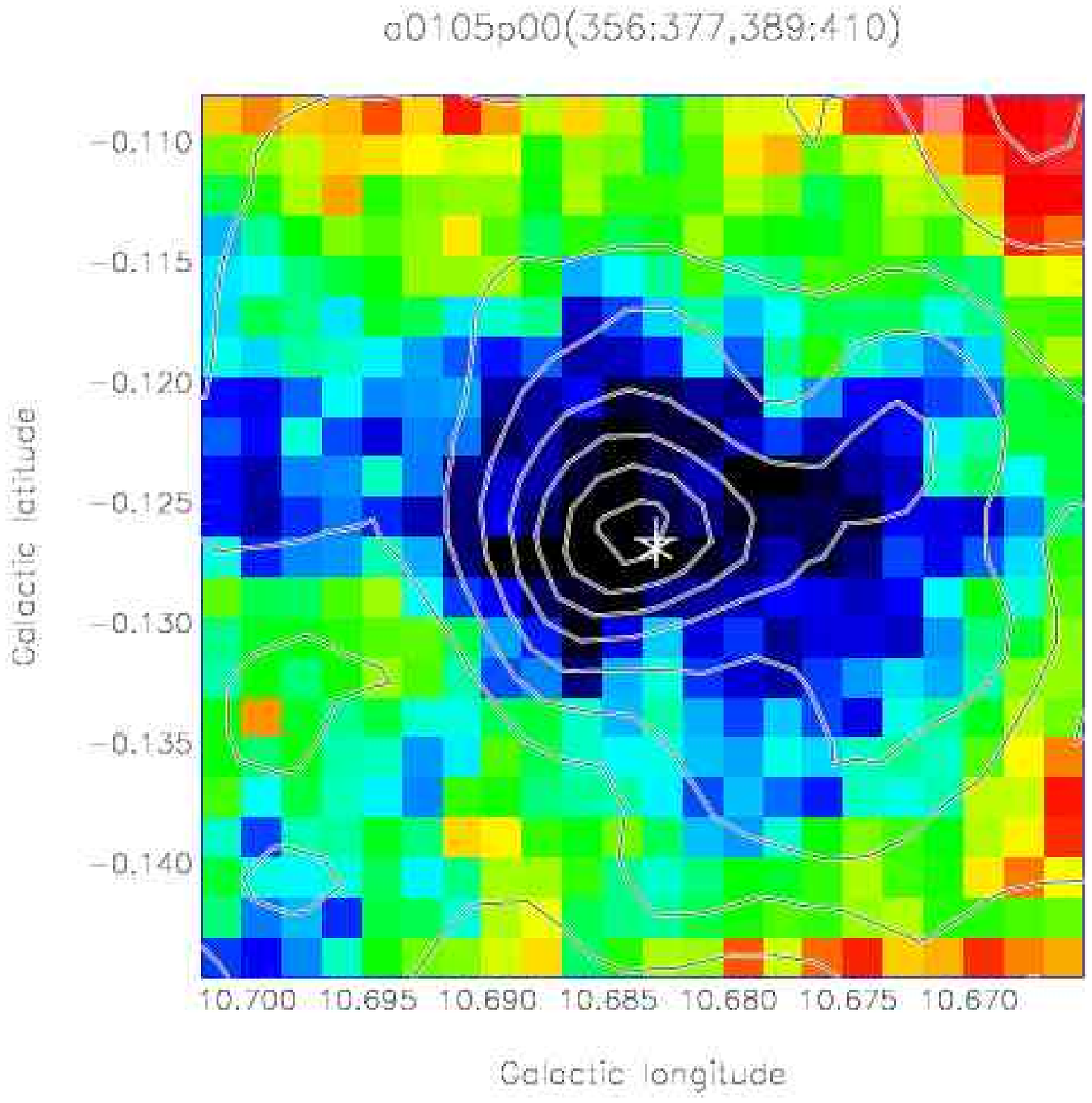}
\includegraphics[width=0.4\textwidth,trim=0 0 0 30,clip]{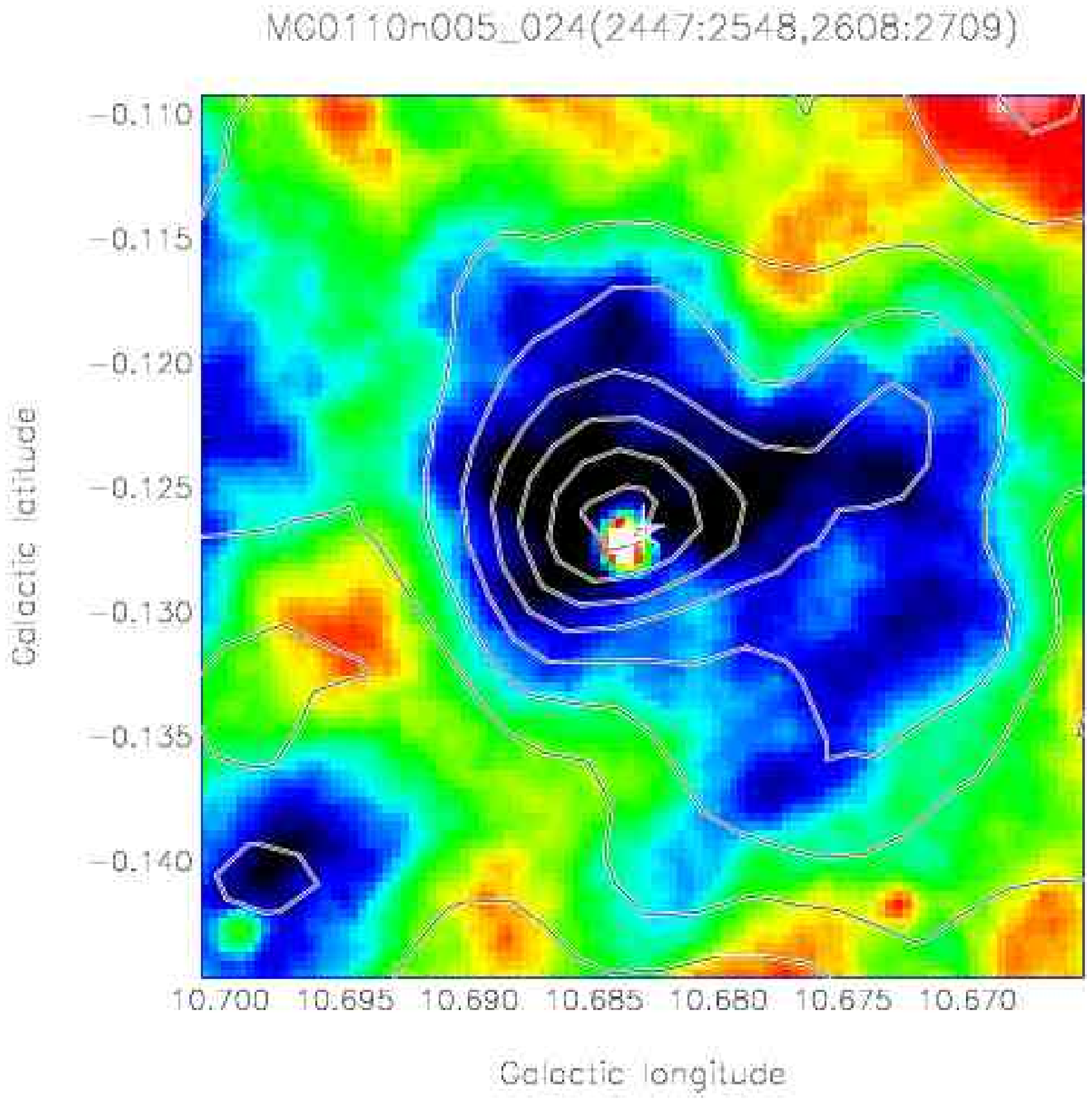}
\includegraphics[width=0.4\textwidth,trim=0 0 0 30,clip]{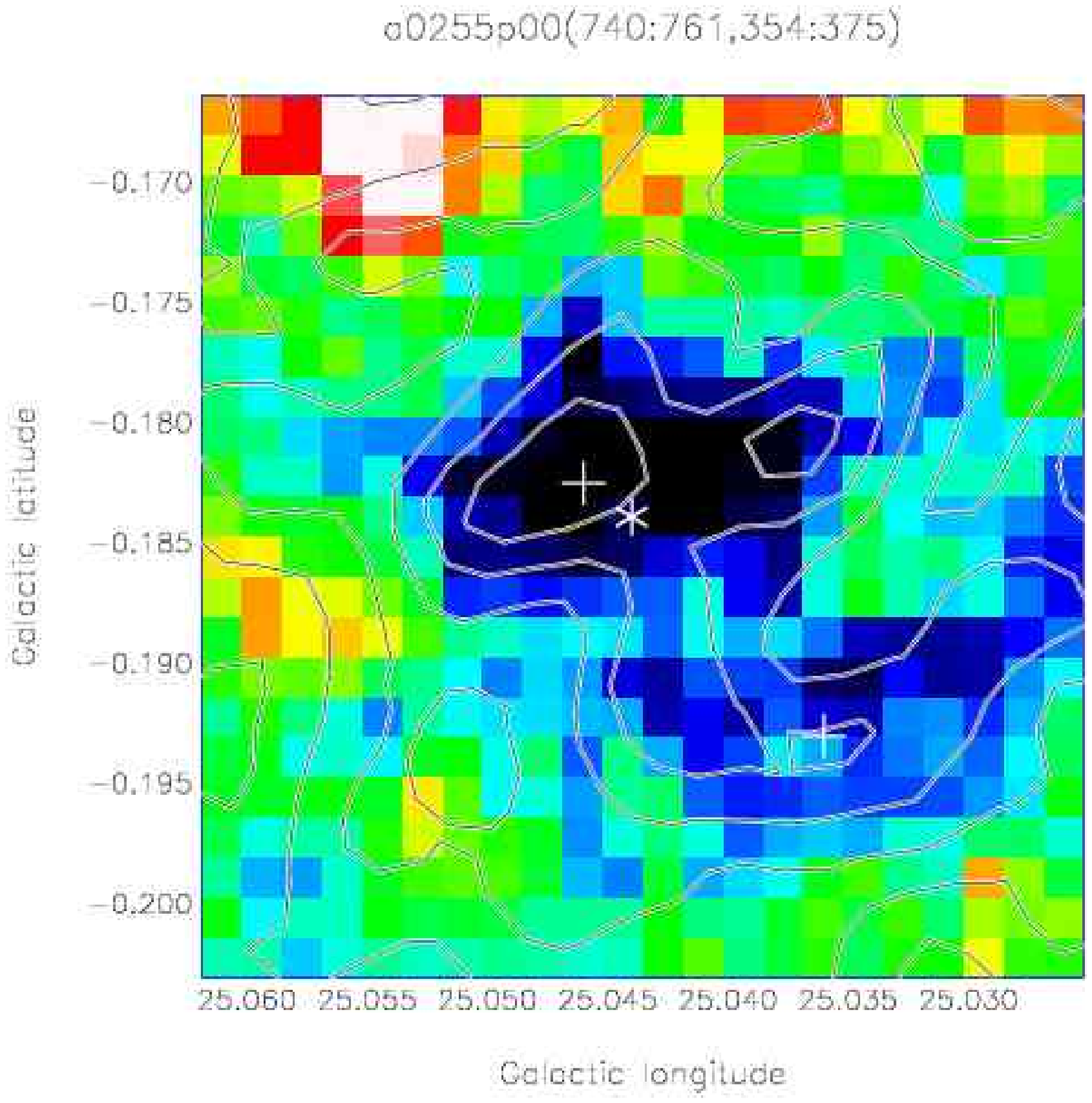}
\includegraphics[width=0.4\textwidth,trim=0 0 0 30,clip]{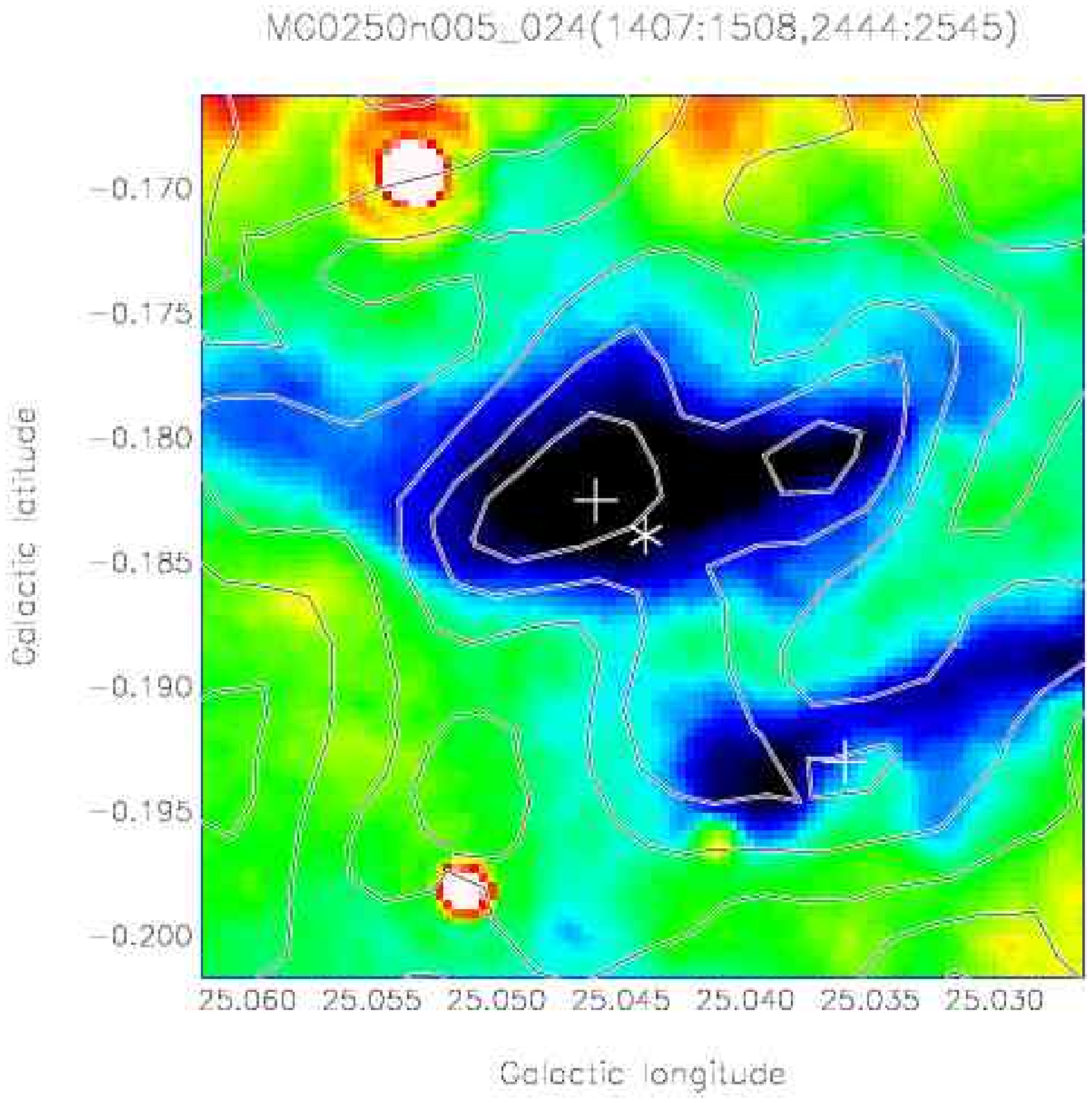}
 \caption{Images of two IRDC cores (top and above) identified in the SCUBA Legacy Catalogue. Images to the left are MSX 8\,$\umu$m and images to the right are MIPSGAL 24\,$\umu$m. In each image a star marks the location of the MSX identified candidate IRDC core and the cross marks location of the SCUBA Clumpfind object. Contours in all cases are SCUBA 850\,$\umu$m. Top: Image of IRDC core: MSXDCG10.71$-$00.16 (b). Bottom: Image of IRDC core: MSXDCG025.04$-$00.20 (f). }
 \label{Fig:cores1}
\end{figure*}

Of the 154 IRDC cores detected at 850\,$\umu$m, 69 were located within the coverage area of MIPSGAL and 34 out of the 51 cores not detected at 850\,$\umu$m were also located within the MIPSGAL survey coverage area. The cores were visually inspected at 24\,$\umu$m and it was found that 48 of the cores detected at 850\,$\umu$m are positionally associated with one or more 24\,$\umu$m MIPSGAL sources (approximately half contain more than one 24\,$\umu$m source). None of the 34 cores not detected at 850\,$\umu$m are found to be associated with any MIPSGAL 24\,$\umu$m sources. Fig. \ref{Fig:cores1} shows two cores that are seen at both 8\,$\umu$m and 24\,$\umu$m, one with an embedded object and one without.


\section{Results}
\label{Section:Results}

\subsection{MSX identified IRDCs in the SCUBA Legacy Catalogue}

In total 205 candidate IRDC MSX cores were found to be within the SCUBA Legacy Catalogue coverage area. 154 candidate cores had detectable emission at 850\,$\umu$m, and 51 candidate cores did not. Of the 154 cores detected at 850\,$\umu$m, we find that they span a range of peak contrast values (0.11\,--\,0.62), column densities and masses, with the peak contrast distribution having a median of 0.32. 8\,$\umu$m column densities range from 0.56$\times 10^{22}$ to 4.21$\times 10^{22}$\,cm$^{-2}$ with a median of 1.7$\times 10^{22}$\,cm$^{-2}$. Mass estimates of these candidates range from 50 to 4,190\,M$_{\odot}$ with a median of 300\,M$_{\odot}$. Peak contrast values for cores detected and not detected at 850\,$\umu$m are given in Tables \ref{table:classA} and \ref{table:classC} respectively. The physical properties of those cores with distance information available (data taken from \cite{2006ApJ...639..227S} and \cite{2008ApJ...680..349J}) are seen in Table \ref{table:mass}.

\begin{figure}
 \includegraphics[width=0.4\textwidth,angle=-90]{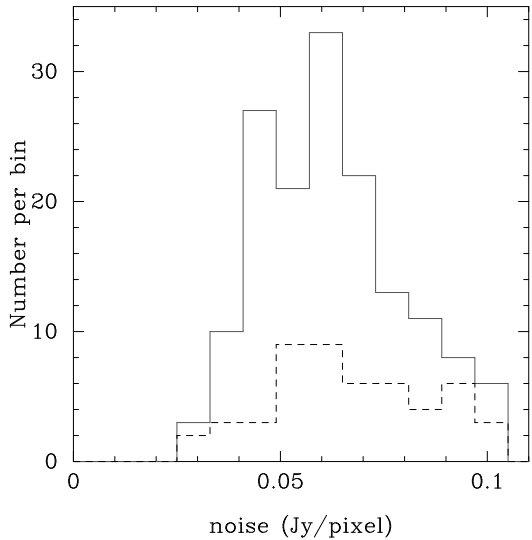}
 \caption{Histogram of rms (Jy/pixel) for cores detected at 850\,$\umu$m (solid line) and cores not detected at 850\,$\umu$m (dashed line). Bin size used is 0.008.}
 \label{Hist:error}
\end{figure}

Those cores not detected at 850\,$\umu$m are found predominantly at low contrast values ($\le$\,0.4), with a mean of 0.22 (seen in Fig. \ref{Hist:ColDen}). This result is not surprising, SCUBA is naturally expected to detect high column density clouds (which would have high contrast values) and not detect low column density clouds (which would have low contrast values). Although the cores not detected at 850\,$\umu$m possess lower peak contrasts, we see no evidence for the existence of two separate populations. A Kolmogorov-Smirnoff (KS) two sample test, on the peak contrast values for cores with and without 850\,$\umu$m emission, reveals no significant difference (to 95\%) that the cores originate from two separate populations. Inspection of the rms values for the 205 candidate IRDC cores, as with the 850\,$\umu$m emission, reveals no significant difference in values between those cores detected at 850\,$\umu$m and those not detected at 850\,$\umu$m as seen in Fig. \ref{Hist:error}. Thus we are confident that the reason behind the cores not being detected at 850\,$\umu$m is not due to them simply lying in high noise regions of the SCUBA Legacy Catalogue. It is possible that some of the cores identified by \citet{2006ApJ...639..227S} that were not detected at 850\,$\umu$m are not true cores at all, rather voids in the mid-infrared background. This possibility is suggested by \cite{2006ApJ...653.1325S} and \cite{2008ApJ...680..349J}, who state at low contrast values the number of mis-identified IRDCs is greater than at high contrasts.

\begin{table*}
 \centering
  \caption{Data for all MSX identified IRDC candidate cores within the SCUBA Legacy Catalogue, with a SCUBA detection. C is the Peak contrast value. F$_{850}$ is the peak flux at 850\,$\umu$m.}
 \begin{minipage}{210mm}
  \begin{tabular}{ccccccccc} 
\hline\hline 
MSX & SCUBA & RA & Dec & C & F$_{850}$ & N$_{8}$(H$_{2}$) & N$_{850}$(H$_{2}$) & MIPS\\
ID & ID & &  &  &  & $\times 10^{22}$ & $\times 10^{22}$ & source \footnote{cores with MIPS data avaliable are denoted by a yes (if an embedded 24\,$\umu$m source is present) or no (if an embedded 24\,$\umu$m is \\not present) in this column}\\
(MSXDC) & (JCMTS\footnote{F stands for Fundamental catalogue and E stands for Extended catalogue}) & (hh:mm:ss) & ( $^{\circ}$: ': '')  & & (Jy/beam) & (cm$^{-2}$) & (cm$^{-2}$) &  \\
\hline
  G000.06+00.21   (a) & F\_J174457.0-284618  & 17:44:55.4 & -28:46:38 & 0.22 & 1.09  & 1.08 & 7.91    &    \\
  G000.07+00.24   (a) & F\_J174450.2-284449  & 17:44:50.7 & -28:44:45 & 0.16 & 0.40  & 0.76 & 2.89    &    \\
  G000.08+00.19   (a) & F\_J174503.4-284618  & 17:45:03.8 & -28:46:09 & 0.15 & 0.76  & 0.77 & 5.51    &    \\
  G000.13$-$00.14 (a) & F\_J174626.0-285437  & 17:46:26.8 & -28:54:38 & 0.22 & 1.41  & 1.08 & 10.22   &    \\
  G000.20$-$00.25 (a) & F\_J174706.6-285314  & 17:47:06.3 & -28:53:26 & 0.19 & 0.57  & 0.92 & 4.13    &    \\
  G000.21+00.23   (a) & F\_J174510.6-283812  & 17:45:13.8 & -28:37:59 & 0.14 & 0.59  & 0.66 & 4.28    &    \\
  G000.25+00.02   (a) & F\_J174607.0-284130  & 17:46:07.9 & -28:41:35 & 0.44 & 6.62  & 2.52 & 47.95   &    \\
  G000.32$-$00.23 (a) & F\_J174719.9-284656  & 17:47:19.3 & -28:47:01 & 0.26 & 0.73  & 1.31 & 5.29    &    \\
  G000.32$-$00.18 (a) & F\_J174707.2-284432  & 17:47:08.6 & -28:45:19 & 0.33 & 0.46  & 1.74 & 3.33    &    \\
  G000.33+00.05   (a) & F\_J174613.0-283654  & 17:46:12.7 & -28:37:10 & 0.21 & 3.74  & 1.02 & 27.21   &    \\
  G000.35$-$00.24 (a) & F\_J174725.4-284538  & 17:47:23.3 & -28:45:47 & 0.16 & 0.34  & 0.76 & 2.46    &    \\
  G000.40$-$00.21 (a) & F\_J174725.4-284156  & 17:47:25.3 & -28:42:16 & 0.23 & 1.26  & 1.14 & 9.14    &    \\
  G000.40+00.04   (a) & F\_J174624.4-283331  & 17:46:24.7 & -28:33:47 & 0.45 & 6.76  & 2.60 & 49.03   &    \\
  G000.59+00.01   (e) & F\_J174646.8-283155  & 17:46:47.5 & -28:32:07 & 0.37 & 14.95 & 2.01 & 112.31  &    \\
  G000.59+00.01   (a) & F\_J174643.2-283007  & 17:46:42.9 & -28:30:27 & 0.47 & 7.83  & 2.76 & 56.80   &    \\
  G000.59+00.01   (c) & F\_J174648.6-282944  & 17:46:49.2 & -28:29:53 & 0.37 & 2.71  & 2.01 & 19.65   &    \\
  G000.59+00.01   (f) & F\_J174653.2-282632  & 17:46:53.0 & -28:26:40 & 0.34 & 2.18  & 1.81 & 15.81   &    \\
  G000.59+00.01   (d) & F\_J174657.8-282332  & 17:46:59.7 & -28:23:06 & 0.37 & 2.10  & 2.01 & 16.20   &    \\
  G000.59+00.01   (b) & F\_J174709.6-282208  & 17:47:09.3 & -28:21:54 & 0.37 & 3.38  & 2.01 & 24.41   &    \\
  G000.68$-$00.18 (a) & F\_J174800.1-282714  & 17:48:01.2 & -28:27:24 & 0.20 & 0.67  & 0.97 & 4.86    &    \\
  G000.72$-$00.08 (a) & F\_J174740.1-282132  & 17:47:39.0 & -28:21:39 & 0.16 & 3.49  & 0.76 & 25.27   &    \\
  G000.87$-$00.01 (a) & E\_J174744.6-281150  & 17:47:45.4 & -28:11:47 & 0.15 & 1.91  & 0.77 & 13.84   &    \\
  G000.90$-$00.02 (a) & E\_J174751.4-281039  & 17:47:51.1 & -28:10:46 & 0.13 & 1.87  & 0.61 & 13.56   &    \\
  G000.94+00.00   (a) & E\_J174750.1-280803  & 17:47:49.5 & -28:07:37 & 0.17 & 1.13  & 0.81 & 8.19    &    \\
  G000.97+00.04   (a) & E\_J174744.6-280445  & 17:47:45.3 & -28:04:45 & 0.22 & 1.18  & 1.08 & 8.55    &    \\
  G000.98+00.09   (a) & E\_J174733.3-280302  & 17:47:32.7 & -28:02:30 & 0.13 & 0.86  & 0.61 & 6.24    &    \\
  G000.97$-$00.06 (c) & E\_J174831.4-280826  & 17:48:30.3 & -28:08:25 & 0.22 & 1.39  & 1.08 & 10.09   &    \\
  G001.01+00.05   (b) & E\_J174752.3-280321  & 17:47:53.0 & -28:03:27 & 0.20 & 0.92  & 0.97 & 8.42    &    \\
  G000.97$-$00.06 (a) & E\_J174814.6-280550  & 17:48:15.3 & -28:06:05 & 0.26 & 1.55  & 1.31 & 11.23   &    \\
  G001.01+00.05   (a) & E\_J174747.8-280157  & 17:47:49.1 & -28:01:58 & 0.20 & 0.90  & 0.97 & 6.52    &    \\
  G001.02+00.02   (a) & E\_J174755.5-280239  & 17:47:57.0 & -28:02:48 & 0.11 & 0.67  & 0.57 & 4.86    &    \\
  G001.11$-$00.16 (a) & F\_J174846.7-280338  & 17:48:47.9 & -28:03:28 & 0.29 & 1.30  & 1.49 & 9.42    &    \\
  G001.11$-$00.16 (b) & F\_J174851.3-280356  & 17:48:52.4 & -28:04:01 & 0.24 & 0.59  & 1.19 & 4.28    &    \\
  G001.13$-$00.26 (a) & F\_J174915.8-280531  & 17:49:16.5 & -28:05:39 & 0.20 & 0.71  & 0.97 & 5.14    &    \\
  G001.26$-$00.23 (a) & F\_J174930.3-275848  & 17:49:31.4 & -27:59:01 & 0.14 & 0.36  & 0.66 & 2.61    &    \\
  G001.26+00.04   (a) & F\_J174826.6-274949  & 17:48:25.3 & -27:49:53 & 0.21 & 1.78  & 1.02 & 12.89   &    \\
  G001.29+00.03   (a) & F\_J174833.4-274836  & 17:48:32.5 & -27:48:45 & 0.17 & 1.62  & 0.81 & 11.75   &    \\
  G001.31$-$00.04 (a) & F\_J174856.9-275047  & 17:48:57.0 & -27:50:28 & 0.17 & 0.67  & 0.81 & 4.86    &    \\
  G001.34$-$00.08 (a) & F\_J174906.9-274946  & 17:49:06.7 & -27:49:48 & 0.13 & 0.42  & 0.61 & 3.05    &    \\
  G001.47$-$00.03 (a) & F\_J174914.3-274211  & 17:49:13.2 & -27:41:55 & 0.18 & 0.40  & 0.86 & 2.89    &    \\
  G001.46+00.03   (a) & E\_J174855.8-273917  & 17:48:56.3 & -27:39:24 & 0.29 & 2.31  & 1.49 & 16.74   &    \\
  G001.51$-$00.10 (a) & F\_J174934.7-274212  & 17:49:36.7 & -27:41:53 & 0.17 & 1.64  & 0.81 & 11.88   &    \\
  G001.53+00.14   (a) & E\_J174842.2-273230  & 17:48:41.0 & -27:33:04 & 0.13 & 0.82  & 0.61 & 5.94    &    \\
  G001.61$-$00.02 (b) & E\_J174936.6-273248  & 17:49:36.5 & -27:33:11 & 0.37 & 0.71  & 2.01 & 5.14    &    \\
  G001.61$-$00.02 (a) & E\_J174944.3-273330  & 17:49:44.4 & -27:33:11 & 0.41 & 2.67  & 2.29 & 19.35   &    \\
  G001.67$-$00.18 (b) & E\_J175015.0-273437  & 17:50:16.0 & -27:34:22 & 0.23 & 0.63  & 1.14 & 4.56    &    \\
  G002.51+00.18   (b) & F\_J175045.8-263945  & 17:50:43.4 & -26:40:21 & 0.38 & 2.77  & 2.08 & 20.09   &    \\
  G002.51+00.18   (a) & F\_J175045.8-263945  & 17:50:48.5 & -26:39:30 & 0.41 & 2.77  & 2.29 & 20.09   &    \\
  G008.83$-$00.05 (e) & F\_J180525.3-211926  & 18:05:27.7 & -21:20:17 & 0.27 & 4.22  & 1.37 & 30.67   &    \\
  G008.83$-$00.05 (a) & F\_J180525.3-211926  & 18:05:26.3 & -21:19:04 & 0.32 & 4.22  & 1.68 & 30.67   &    \\
  G009.84$-$00.03 (a) & E\_J180733.9-202613  & 18:07:37.4 & -20:26:20 & 0.43 & 0.65  & 2.44 & 4.71    &  yes  \\
  G010.71$-$00.16 (g) & E\_J180944.4-194712  & 18:09:44.4 & -19:47:02 & 0.35 & 0.46  & 1.87 & 3.33    &  yes  \\
  G010.71$-$00.16 (h) & E\_J180953.3-194806  & 18:09:52.5 & -19:47:40 & 0.34 & 0.65  & 1.81 & 4.71    &  yes  \\
 \hline\hline
\label{table:classA}
\end{tabular}
\end{minipage}
\end{table*}

\begin{table*}
\setcounter{table}{0}
 \centering
  \caption{- Continued}
 \begin{minipage}{210mm}
  \begin{tabular}{ccccccccc} 
\hline\hline 
MSX & SCUBA & RA & Dec & C & F$_{850}$ & N$_{8}$(H$_{2}$) & N$_{850}$(H$_{2}$) & MIPS\\
ID & ID & &  &  &  & $\times 10^{22}$ & $\times 10^{22}$ & source \\
(MSXDC) & (JCMTS) & (hh:mm:ss) & ( $^{\circ}$: ': '')  & & (Jy/beam) & (cm$^{-2}$) & (cm$^{-2}$) &  \\
\hline
  G010.71$-$00.16 (b) & E\_J180938.7-194512  & 18:09:38.7 & -19:45:15 & 0.49 & 2.79  & 2.93 & 20.22   &  yes  \\
  G010.71$-$00.16 (d) & E\_J180949.1-194442  & 18:09:48.9 & -19:44:52 & 0.40 & 0.44  & 2.22 & 3.20    &  no   \\
  G010.71$-$00.16 (f) & E\_J180940.8-194336  & 18:09:41.5 & -19:43:42 & 0.36 & 2.35  & 1.94 & 17.04   &  no   \\
  G010.71$-$00.16 (a) & E\_J180945.7-194206  & 18:09:45.5 & -19:42:22 & 0.56 & 1.28  & 3.57 & 9.29    &  yes  \\
  G010.71$-$00.16 (e) & E\_J181009.0-194507  & 18:10:10.0 & -19:45:13 & 0.38 & 0.63  & 2.80 & 4.56    &  no   \\
  G010.71$-$00.16 (c) & E\_J181003.1-194342  & 18:10:02.4 & -19:43:28 & 0.44 & 0.84  & 2.52 & 6.09    &  yes  \\
  G010.99$-$00.07 (a) & F\_J181007.1-192755  & 18:10:07.2 & -19:27:60 & 0.55 & 1.39  & 3.47 & 10.09   &  yes  \\
  G011.11$-$00.11 (g) & F\_J181013.5-192419  & 18:10:13.7 & -19:24:36 & 0.30 & 0.67  & 1.55 & 4.86    &  no   \\
  G011.11$-$00.11 (b) & F\_J181018.2-192431  & 18:10:18.7 & -19:24:42 & 0.43 & 0.88  & 2.44 & 6.37    &  yes  \\
  G011.11$-$00.11 (a) & F\_J181033.0-192201  & 18:10:32.0 & -19:22:31 & 0.50 & 1.45  & 3.01 & 10.52   &  yes  \\
  G011.11$-$00.11 (f) & F\_J181033.0-192201  & 18:10:35.4 & -19:21:02 & 0.30 & 0.73  & 1.55 & 5.29    &  yes  \\
  G011.11$-$00.11 (d) & F\_J181037.3-191820  & 18:10:37.7 & -19:18:27 & 0.31 & 0.36  & 1.61 & 2.61    &  yes  \\
  G011.11$-$00.11 (c) & F\_J181034.8-191702  & 18:10:35.3 & -19:17:20 & 0.40 & 0.80  & 2.22 & 5.81    &  yes  \\
  G011.11$-$00.11 (e) & F\_J181039.9-191132  & 18:10:40.6 & -19:10:58 & 0.31 & 0.76  & 1.61 & 5.51    &  yes  \\
  G012.44$-$00.14 (b) & F\_J181320.6-181220  & 18:13:21.6 & -18:12:17 & 0.34 & 0.36  & 1.81 & 2.61    &  yes  \\
  G012.44$-$00.14 (a) & F\_J181341.7-181239  & 18:13:41.5 & -18:12:32 & 0.45 & 1.43  & 2.60 & 20.37   &  yes  \\
  G012.44$-$00.14 (d) & F\_J181331.6-181115  & 18:13:32.5 & -18:11:18 & 0.29 & 0.36  & 1.49 & 2.61    &  no   \\
  G012.88+00.53   (a) & F\_J181145.3-173044  & 18:11:44.8 & -17:31:17 & 0.36 & 1.11  & 1.94 & 8.06    &  yes  \\
  G013.02$-$00.83 (a) & F\_J181700.1-180202  & 18:17:00.7 & -18:02:18 & 0.47 & 0.27  & 2.76 & 1.97    &  yes  \\
  G013.68$-$00.60 (a) & F\_J181725.7-172049  & 18:17:27.6 & -17:20:59 & 0.24 & 0.42  & 1.19 & 3.05    &  no   \\
  G018.50$-$00.16 (c) & E\_J182517.5-125526  & 18:25:17.5 & -12:55:31 & 0.27 & 0.29  & 1.37 & 2.10    &  no   \\
  G018.50$-$00.16 (d) & F\_J182523.2-125450  & 18:25:23.3 & -12:54:55 & 0.25 & 0.59  & 1.25 & 4.28    &  no   \\
  G018.50$-$00.16 (b) & F\_J182520.4-125014  & 18:25:21.2 & -12:50:18 & 0.29 & 0.32  & 1.49 & 2.31    &  no   \\
  G018.58$-$00.08 (b) & F\_J182507.3-124750  & 18:25:07.6 & -12:48:00 & 0.32 & 0.46  & 1.68 & 3.33    &  yes  \\
  G018.58$-$00.08 (a) & F\_J182508.5-124520  & 18:25:08.9 & -12:45:20 & 0.37 & 1.91  & 2.01 & 13.84   &  yes  \\
  G019.27+00.07   (a) & F\_J182552.1-120456  & 18:25:54.0 & -12:04:56 & 0.50 & 1.03  & 3.01 & 7.47    &  yes  \\
  G022.35+00.41   (b) & F\_J183029.6-091238  & 18:30:28.7 & -09:12:31 & 0.37 & 0.38  & 2.01 & 2.76    &  no   \\
  G022.35+00.41   (a) & F\_J183024.4-091038  & 18:30:24.7 & -09:10:47 & 0.51 & 1.89  & 3.10 & 13.69   &  yes  \\
  G023.86$-$00.19 (a) & E\_J183526.9-080814  & 18:35:26.6 & -08:08:22 & 0.32 & 0.42  & 1.68 & 3.05    &  no   \\
  G024.00+00.15   (a) & E\_J183428.8-075220  & 18:34:29.5 & -07:52:23 & 0.20 & 1.11  & 0.97 & 8.06    &  yes  \\
  G024.36$-$00.16 (a) & F\_J183618.3-074102  & 18:36:17.5 & -07:41:27 & 0.41 & 0.88  & 2.29 & 6.37    &  yes  \\
  G024.37$-$00.21 (a) & F\_J183630.0-074208  & 18:36:30.2 & -07:42:16 & 0.34 & 0.34  & 1.81 & 2.46    &  yes  \\
  G024.60+00.08   (a) & F\_J183540.1-071838  & 18:35:39.4 & -07:18:51 & 0.49 & 2.02  & 2.93 & 14.64   &  yes  \\
  G024.68+00.17   (a) & F\_J183540.1-071514  & 18:35:41.2 & -07:15:22 & 0.20 & 0.25  & 0.97 & 1.81    &  yes  \\
  G025.04$-$00.20 (g) & F\_J183712.0-071126  & 18:37:12.8 & -07:11:23 & 0.36 & 0.76  & 1.94 & 5.51    &  yes  \\
  G025.04$-$00.20 (e) & F\_J183719.2-071144  & 18:37:18.8 & -07:11:49 & 0.41 & 1.13  & 2.29 & 8.19    &  yes  \\
  G025.04-00.20   (b) &  F\_J183734.6-070726 & 18:37:34.8 & -07:07:39 & 0.44 & 0.38  & 2.52 & 2.76    &  yes  \\
  G025.04$-$00.20 (f) & F\_J183738.2-070550  & 18:37:38.2 & -07:06:00 & 0.38 & 0.29  & 2.08 & 2.10    &  no   \\
  G028.37+00.07   (a) & F\_J184250.6-040314  & 18:42:50.6 & -04:03:30 & 0.61 & 2.52  & 4.09 & 18.27   &  yes  \\
  G028.37+00.07   (d) & F\_J184248.2-040133  & 18:42:48.6 & -04:01:42 & 0.47 & 0.65  & 2.76 & 4.71    &  yes  \\
  G028.37+00.07   (b) & F\_J184255.4-040150  & 18:42:55.5 & -04:01:47 & 0.51 & 0.95  & 3.10 & 6.89    &  no   \\
  G028.37+00.07   (e) & F\_J184300.2-040132  & 18:43:00.5 & -04:01:36 & 0.45 & 0.69  & 2.60 & 5.01    &  no   \\
  G028.53$-$00.25 (g) & F\_J184417.3-040208  & 18:44:17.0 & -04:02:18 & 0.27 & 0.61  & 1.37 & 5.40    &  yes  \\
  G028.53$-$00.25 (b) & F\_J184422.5-040150  & 18:44:23.7 & -04:02:09 & 0.38 & 0.57  & 2.08 & 4.13    &  yes  \\
  G028.53$-$00.25 (c) & F\_J184415.6-040056  & 18:44:16.6 & -04:01:02 & 0.34 & 1.03  & 1.81 & 7.47    &  yes  \\
  G028.53$-$00.25 (a) & F\_J184418.1-035938  & 18:44:17.1 & -03:59:37 & 0.41 & 2.65  & 2.29 & 19.20   &  yes  \\
  G028.53$-$00.25 (e) & F\_J184418.1-035938  & 18:44:17.7 & -03:58:16 & 0.29 & 0.69  & 1.49 & 5.01    &  yes  \\
  G028.61$-$00.26 (a) & F\_J184428.1-035750  & 18:44:29.0 & -03:57:46 & 0.27 & 0.32  & 1.37 & 2.31    &  yes  \\
  G030.77+00.22   (a) & F\_J184647.8-014856  & 18:46:47.1 & -01:49:03 & 0.25 & 4.62  & 1.25 & 33.48   &  yes  \\
  G030.97$-$00.14 (a) & E\_J184821.9-014832  & 18:48:24.2 & -01:48:25 & 0.38 & 1.89  & 2.08 & 13.69   &  yes  \\
  G031.03+00.26   (b) & F\_J184701.4-013438  & 18:47:01.5 & -01:34:47 & 0.29 & 0.78  & 1.49 & 5.66    &  yes  \\
  G031.03+00.26   (c) & F\_J184707.4-013432  & 18:47:07.7 & -01:34:42 & 0.29 & 0.50  & 1.49 & 3.63    &  yes  \\
  G031.03+00.26   (a) & F\_J184701.4-013314  & 18:47:01.2 & -01:33:23 & 0.31 & 0.55  & 1.61 & 4.00    &  yes  \\
  G031.23+00.05   (a) & E\_J184807.5-012844  & 18:48:08.3 & -01:28:50 & 0.21 & 0.82  & 1.02 & 5.94    &  yes  \\
  G031.27+00.08   (a) & E\_J184807.9-012626  & 18:48:08.0 & -01:26:43 & 0.16 & 0.69  & 0.76 & 5.01    &  no   \\
  G031.38+00.29   (a) & F\_J184732.6-011338  & 18:47:34.0 & -01:13:59 & 0.38 & 1.53  & 2.08 & 11.10   &  no   \\
  G031.97+00.07   (b) & F\_J184922.1-005038  & 18:49:22.2 & -00:50:47 & 0.42 & 0.59  & 2.37 & 4.28    &  yes  \\
  G031.97+00.07   (c) & F\_J184926.9-005002  & 18:49:27.0 & -00:50:11 & 0.35 & 0.36  & 1.87 & 2.61    &  no   \\
  G033.69$-$00.01 (e) & E\_J185248.6+003602  & 18:52:49.6 & +00:35:55 & 0.32 & 0.63  & 1.68 & 4.56    &  no   \\
  G033.69$-$00.01 (b) & E\_J185252.6+003832  & 18:52:53.1 & +00:38:10 & 0.38 & 0.65  & 2.08 & 4.71    &  no   \\
  G033.69$-$00.01 (c) & E\_J185253.8+004120  & 18:52:53.0 & +00:40:44 & 0.37 & 0.59  & 2.01 & 4.28    &  no   \\
 \hline\hline
\label{table:classA2}
\end{tabular}
\end{minipage}
\end{table*}

\begin{table*}
\setcounter{table}{0}
 \centering
  \caption{- Continued}
 \begin{minipage}{210mm}
  \begin{tabular}{ccccccccc} 
\hline\hline 
MSX & SCUBA & RA & Dec & C & F$_{850}$ & N$_{8}$(H$_{2}$) & N$_{850}$(H$_{2}$) & MIPS\\
ID & ID & &  &  &  & $\times 10^{22}$ & $\times 10^{22}$ & source \\
(MSXDC) & (JCMTS) & (hh:mm:ss) & ( $^{\circ}$: ': '')  & & (Jy/beam) & (cm$^{-2}$) & (cm$^{-2}$) &  \\
\hline
  G033.69$-$00.01 (a) & E\_J185257.0+004302  & 18:52:57.6 & +00:42:59 & 0.38 & 1.60  & 2.08 & 11.60   &  yes  \\
  G034.43+00.24   (a) & F\_J185318.9+012650  & 18:53:18.9 & +01:26:39 & 0.34 & 0.38  & 1.81 & 2.76    &  yes  \\
  G038.95$-$00.47 (a) & F\_J190407.5+050844  & 19:04:08.3 & +05:08:49 & 0.53 & 1.37  & 3.28 & 9.94    &  yes  \\
  G048.52$-$00.47 (a) & F\_J192207.4+133713  & 19:22:07.9 & +13:36:58 & 0.38 & 0.29  & 2.08 & 2.10    &  no   \\
  G048.65$-$00.29 (a) & F\_J192144.7+134925  & 19:21:45.3 & +13:49:22 & 0.34 & 0.40  & 1.81 & 2.89    &  yes  \\
  G079.24+00.41   (b) & E\_J203137.7+401935  & 20:31:38.1 & +40:19:38 & 0.51 & 0.88  & 3.10 & 6.37    &       \\
  G079.24+00.41   (a) & F\_J203157.6+401828  & 20:31:56.8 & +40:18:12 & 0.52 & 1.99  & 3.19 & 14.43   &       \\
  G081.67+00.44   (a) & F\_J203924.9+421555  & 20:39:21.0 & +42:15:47 & 0.23 & 2.31  & 1.14 & 16.74   &       \\
  G081.73+00.59   (a) & F\_J203859.3+422330  & 20:38:58.2 & +42:23:55 & 0.27 & 8.19  & 1.37 & 59.39   &       \\
  G081.76+00.63   (a) & F\_J203851.6+422717  & 20:38:52.6 & +42:27:12 & 0.17 & 0.80  & 0.81 & 5.81    &       \\
  G351.52+00.69   (a) & F\_J172056.6-354044  & 17:20:58.2 & -35:40:28 & 0.39 & 2.48  & 2.15 & 17.97   &       \\
  G353.26$-$00.16 (a) & F\_J172935.1-344316  & 17:29:33.6 & -34:43:31 & 0.31 & 0.44  & 1.61 & 3.20    &       \\
  G353.37$-$00.33 (b) & F\_J173012.1-344328  & 17:30:12.4 & -34:43:45 & 0.38 & 1.47  & 2.08 & 10.65   &       \\
  G353.37$-$00.33 (a) & F\_J173017.0-344217  & 17:30:18.8 & -34:41:58 & 0.43 & 2.29  & 2.44 & 19.61   &       \\
  G353.90+00.25   (e) & F\_J172902.5-335950  & 17:29:02.2 & -34:00:12 & 0.46 & 1.13  & 2.68 & 8.19    &       \\
  G353.90+00.25   (a) & F\_J172917.1-340017  & 17:29:12.8 & -34:00:01 & 0.62 & 0.55  & 4.21 & 4.00    &       \\
  G353.90+00.25   (f) & F\_J172917.1-340017  & 17:29:16.8 & -34:00:25 & 0.46 & 0.61  & 2.68 & 5.40    &       \\
  G353.90+00.25   (c) & F\_J172919.4-335550  & 17:29:19.1 & -33:55:59 & 0.50 & 0.59  & 3.01 & 4.28    &       \\
  G353.90+00.25   (b) & F\_J172928.5-335444  & 17:29:27.9 & -33:55:06 & 0.59 & 0.50  & 3.88 & 3.63    &       \\
  G353.90+00.25   (d) & F\_J172927.1-335302  & 17:29:25.0 & -33:53:03 & 0.49 & 0.90  & 2.93 & 6.52    &       \\
  G359.05+00.00   (a) & F\_J174321.6-294437  & 17:43:21.8 & -29:44:43 & 0.27 & 0.63  & 1.37 & 4.56    &       \\
  G359.06$-$00.03 (a) & F\_J174326.7-294531  & 17:43:29.7 & -29:45:22 & 0.16 & 0.71  & 0.76 & 5.14    &       \\
  G359.08+00.04   (a) & F\_J174314.8-294143  & 17:43:14.8 & -29:42:02 & 0.20 & 0.50  & 0.97 & 3.63    &       \\
  G359.29$-$00.03 (a) & F\_J174404.5-293302  & 17:44:03.5 & -29:33:12 & 0.21 & 1.28  & 1.02 & 9.29    &       \\
  G359.37$-$00.28 (a) & F\_J174514.5-293644  & 17:45:14.1 & -29:37:27 & 0.12 & 0.27  & 0.56 & 19.57   &       \\
  G359.41+00.08   (a) & F\_J174354.0-292314  & 17:43:53.9 & -29:23:27 & 0.16 & 1.03  & 0.76 & 7.47    &       \\
  G359.46$-$00.03 (a) & F\_J174428.9-292426  & 17:44:29.3 & -29:24:18 & 0.28 & 2.42  & 1.43 & 19.54   &       \\
  G359.48$-$00.22 (a) & F\_J174514.4-292902  & 17:45:15.1 & -29:29:30 & 0.14 & 1.87  & 0.66 & 13.56   &       \\
  G359.59+00.02   (a) & F\_J174436.3-291621  & 17:45:33.5 & -29:24:26 & 0.31 & 1.87  & 1.61 & 15.56   &       \\
  G359.60$-$00.22 (b) & F\_J174535.0-292456  & 17:44:32.0 & -29:16:05 & 0.20 & 1.97  & 0.97 & 14.28   &       \\
  G359.60$-$00.22 (a) & F\_J174535.0-292314  & 17:45:35.2 & -29:23:15 & 0.36 & 5.50  & 1.94 & 39.96   &       \\
  G359.68$-$00.13 (a) & F\_J174526.3-291608  & 17:45:25.3 & -29:16:06 & 0.20 & 0.61  & 0.97 & 5.40    &       \\
  G359.80$-$00.13 (a) & F\_J174539.0-291132  & 17:45:37.1 & -29:11:25 & 0.32 & 1.62  & 1.68 & 11.75   &       \\
  G359.82+00.12   (a) & F\_J174443.6-290127  & 17:44:43.1 & -29:01:02 & 0.19 & 0.44  & 0.92 & 3.20    &       \\
  G359.83+00.18   (a) & E\_J174429.3-285901  & 17:44:29.5 & -28:59:13 & 0.21 & 0.38  & 1.02 & 2.76    &       \\
  G359.85+00.21   (a) & E\_J174425.2-285643  & 17:44:25.3 & -28:56:49 & 0.23 & 0.38  & 1.14 & 2.76    &       \\
  G359.87$-$00.09 (a) & F\_J174544.0-290502  & 17:45:43.4 & -29:05:22 & 0.25 & 5.44  & 1.25 & 3.95    &       \\
  G359.90$-$00.30 (a) & F\_J174636.3-291011  & 17:46:35.8 & -29:10:26 & 0.35 & 2.33  & 1.87 & 16.89   &       \\
  G359.91+00.17   (b) & F\_J174444.4-285519  & 17:44:43.6 & -28:55:28 & 0.45 & 0.90  & 2.60 & 6.52    &       \\
  G359.91+00.17   (a) & F\_J174448.5-285349  & 17:44:47.6 & -28:53:54 & 0.58 & 4.41  & 3.77 & 31.96   &       \\
\hline\hline
\label{table:classA3}
\end{tabular}
\end{minipage}
\end{table*}

\begin{table*} 
 \centering
  \caption{Data for all MSX identified IRDC candidate cores within the SCUBA Legacy Catalogue, with no detection at 850\,$\umu$m.}
 \begin{minipage}{140mm}
  \begin{tabular}{ccccccc} 
\hline\hline
MSX ID &  l & b & RA & Dec & Peak & N$_{8}$(H$_{2}$) \\
(MSXDC)  & ($^{\circ}$) & ($^{\circ}$) & (hh:mm:ss) & ( $^{\circ}$: ': '') & Contrast & $\times 10^{22}$\,cm$^{-2}$ \\
\hline
  G000.13$-$00.14 (b)  & 0.163   & -0.164  & 17:46:38.8  & -28:52:56 &  0.19   & 0.92    \\
  G000.13$-$00.19 (a)  & 0.131   & -0.196  & 17:46:41.8  & -28:55:34 &  0.21   & 1.02    \\
  G000.36$-$00.21 (a)  & 0.364   & -0.216  & 17:47:19.6  & -28:44:14 &  0.20   & 0.97    \\
  G000.73$-$00.01 (a)  & 0.733   & -0.014  & 17:47:24.6  & -28:19:02 &  0.15   & 0.77    \\
  G001.62$-$00.08 (a)  & 1.631   & -0.092  & 17:49:48.6  & -27:35:17 &  0.16   & 0.76    \\
  G004.33$-$00.04 (a)  & 4.336   & -0.052  & 17:55:48.0  & -25:14:19 &  0.34   & 1.81    \\
  G004.33$-$00.04 (c)  & 4.358   & -0.056  & 17:55:51.9  & -25:13:18 &  0.29   & 1.49    \\
  G006.06-01.39   (a)  & 6.063   & -1.397  & 18:04:43.5  & -24:24:28 &  0.23   & 1.14    \\
  G006.09-01.39   (a)  & 6.094   & -1.396  & 18:04:47.3  & -24:22:49 &  0.24   & 1.19    \\
  G006.09-01.36   (a)  & 6.094   & -1.367  & 18:04:40.6  & -24:21:58 &  0.20   & 0.97    \\
  G010.57$-$00.30 (a)  & 10.588  & -0.311  & 18:10:08.1  & -19:55:35 &  0.33   & 1.74    \\
  G010.94$-$00.05 (a)  & 10.944  & -0.059  & 18:09:55.7  & -19:29:35 &  0.18   & 0.86    \\
  G012.37+00.50   (c)  & 12.431  & 0.496   & 18:10:54.4  & -17:55:22 &  0.29   & 1.49    \\
  G012.44$-$00.20 (a)  & 12.449  & -0.201  & 18:13:30.9  & -18:14:31 &  0.15   & 0.77    \\
  G012.88+00.53   (c)  & 12.886  & 0.528   & 18:11:42.5  & -17:30:31 &  0.29   & 1.49    \\
  G013.15+00.09   (a)  & 13.154  & 0.099   & 18:13:49.6  & -17:28:46 &  0.25   & 1.25    \\
  G017.00+00.67   (a)  & 17.003  & 0.661   & 18:19:22.5  & -13:49:34 &  0.33   & 1.74    \\
  G017.01+00.78   (a)  & 17.013  & 0.789   & 18:18:55.8  & -13:45:24 &  0.13   & 0.61    \\
  G017.03+00.71   (a)  & 17.029  & 0.719   & 18:19:13.0  & -13:46:33 &  0.14   & 0.66    \\
  G017.10+00.71   (a)  & 17.096  & 0.699   & 18:19:25.1  & -13:43:34 &  0.31   & 1.61    \\
  G017.10+00.71   (b)  & 17.129  & 0.711   & 18:19:26.4  & -13:41:29 &  0.31   & 1.61    \\
  G018.99$-$00.30 (b)  & 19.004  & -0.306  & 18:26:44.4  & -12:30:45 &  0.28   & 1.43    \\
  G024.60+00.08   (b)  & 24.659  & 0.163   & 18:35:40.9  & -07:16:57 &  0.32   & 1.68    \\
  G024.60+00.08   (d)  & 24.596  & 0.131   & 18:35:40.7  & -07:21:11 &  0.29   & 1.49    \\
  G025.12$-$00.16 (a)  & 25.126  & -0.162  & 18:37:42.6  & -07:01:01 &  0.12   & 0.56    \\
  G025.37$-$00.06 (a)  & 25.419  & -0.104  & 18:38:02.7  & -06:43:48 &  0.26   & 1.31    \\
  G025.42+00.10   (a)  & 25.426  & 0.109   & 18:37:17.7  & -06:37:34 &  0.14   & 0.66    \\
  G027.93$-$00.34 (a)  & 27.924  & -0.344  & 18:43:30.8  & -04:36:47 &  0.19   & 0.92    \\
  G031.32+00.09   (a)  & 31.326  & 0.094   & 18:48:10.4  & -01:23:11 &  0.17   & 0.81    \\
  G031.33+00.12   (a)  & 31.336  & 0.124   & 18:48:05.1  & -01:21:49 &  0.16   & 0.76    \\
  G033.69$-$00.01 (d)  & 33.639  & -0.056  & 18:52:55.6  & +00:36:14 &  0.35   & 1.87    \\
  G034.94+00.37   (a)  & 34.941  & 0.384   & 18:53:44.2  & +01:57:47 &  0.18   & 0.86    \\
  G042.75+00.01   (a)  & 42.751  & 0.019   & 19:09:25.2  & +08:44:19 &  0.21   & 1.02    \\
  G042.75$-$00.19 (a)  & 42.753  & -0.202  & 19:10:13.0  & +08:38:18 &  0.22   & 1.08    \\
  G080.88$-$00.12 (a)  & 80.886  & -0.131  & 20:39:16.7  & +41:17:23 &  0.21   & 1.02    \\
  G081.49+00.13   (a)  & 81.504  & 0.129   & 20:40:11.5  & +41:56:17 &  0.40   & 2.22    \\
  G081.49+00.13   (b)  & 81.498  & 0.161   & 20:40:02.2  & +41:57:11 &  0.34   & 1.81    \\
  G081.56+00.57   (a)  & 81.564  & 0.581   & 20:38:27.4  & +42:15:40 &  0.16   & 0.76    \\
  G081.57+00.50   (b)  & 81.576  & 0.523   & 20:38:44.7  & +42:14:07 &  0.27   & 1.37    \\
  G081.60+00.58   (a)  & 81.603  & 0.586   & 20:38:33.7  & +42:17:42 &  0.17   & 0.81    \\
  G081.69+00.71   (a)  & 81.699  & 0.708   & 20:38:21.2  & +42:26:43 &  0.19   & 0.92    \\
  G351.50+00.66   (a)  & 351.509 & 0.661   & 17:21:05.1  & -35:41:56 &  0.17   & 0.81    \\
  G353.26$-$00.16 (f)  & 353.281 & -0.207  & 17:29:28.1  & -34:43:09 &  0.24   & 1.19    \\
  G353.90+00.25   (g)  & 353.886 & 0.254   & 17:29:13.6  & -33:57:36 &  0.40   & 2.22    \\
  G353.98+00.39   (a)  & 353.993 & 0.391   & 17:28:57.7  & -33:47:43 &  0.20   & 0.97    \\
  G359.25+00.01   (a)  & 359.254 & 0.016   & 17:43:46.2  & -29:33:50 &  0.15   & 0.77    \\
  G359.28+00.02   (a)  & 359.298 & 0.031   & 17:43:49.0  & -29:31:06 &  0.38   & 2.08    \\
  G359.28+00.02   (b)  & 359.298 & 0.006   & 17:43:54.9  & -29:31:54 &  0.34   & 1.81    \\
  G359.79$-$00.25 (b)  & 359.799 & -0.267  & 17:46:11.1  & -29:14:48 &  0.23   & 1.14    \\
  G359.81$-$00.29 (a)  & 359.814 & -0.297  & 17:46:20.3  & -29:14:58 &  0.17   & 0.81    \\
  G359.82$-$00.37 (b)  & 359.843 & -0.367  & 17:46:40.9  & -29:15:40 &  0.27   & 1.37    \\
\hline\hline
\label{table:classC}
\end{tabular}
\end{minipage}
\end{table*}

\begin{table*}
 \centering
  \caption{Mass estimates for clouds with distances obtained by the GRS survey \citep{2008ApJ...680..349J}. Galacitc coordinates quoted come from the MSX catalogue positions \citep{2006ApJ...639..227S}.}
  \begin{minipage}{200mm}
  \begin{tabular}{ccccccccccc} 
\hline\hline
MSX ID & SCUBA ID & l & b  & Peak & Flux \footnote{Flux integrated over the area of the object as defined by Clumpfind \cite{2008ApJS..175..277D}} & Distance & Mass & MIPS \\ 
(MSXDC)  & (JCMTS) & ($^{\circ}$) & ($^{\circ}$) & Contrast & Jy & kpc & M$_{\odot}$ & detected? \\  
\hline 
G018.50$-$00.16 (b) & F\_J182520.4-125014  & 18.558  & -0.159  & 0.29  & 1.79  & 4.1  &  240  & No  \\ 
G018.58$-$00.08 (b) & F\_J182507.3-124750  & 18.566  & -0.092  & 0.32  & 0.97  & 3.8  &  110  & Yes \\ 
G018.58$-$00.08 (a) & F\_J182508.5-124520  & 18.608  & -0.076  & 0.37  & 4.51  & 3.8  &  500  & Yes \\ 
G019.27+00.07   (a) & F\_J182552.1-120456  & 19.289  & 0.076   & 0.50  & 3.24  & 2.4  &  130  & Yes \\ 
G022.35+00.41   (b) & F\_J183029.6-091238  & 22.356  & 0.416   & 0.37  & 2.02  & 4.3  &  290  & No  \\ 
G022.35+00.41   (a) & F\_J183024.4-091038  & 22.374  & 0.444   & 0.51  & 4.94  & 4.3  &  720  & Yes \\ 
G023.86$-$00.19 (a) & E\_J183526.9-080814  & 23.871  & -0.179  & 0.32  & 1.92  & 4.0  &  240  & No  \\ 
G024.36$-$00.16 (a) & F\_J183618.3-074102  & 24.366  & -0.159  & 0.41  & 1.77  & 3.9  &  190  & Yes \\ 
G024.37$-$00.21 (a) & F\_J183630.0-074208  & 24.378  & -0.212  & 0.34  & 0.46  & 3.9  &  50   & Yes \\ 
G024.60+00.08   (a) & F\_J183540.1-071838  & 24.628  & 0.154   & 0.49  & 5.22  & 3.7  &  560  & Yes \\ 
G028.37+00.07   (a) & F\_J184250.6-040314  & 28.341  & 0.058   & 0.61  & 10.8  & 5.0  &  1010 & Yes \\ 
G028.37+00.07   (d) & F\_J184248.2-040133  & 28.364  & 0.079   & 0.47  & 2.52  & 5.0  &  100  & Yes \\ 
G028.37+00.07   (g) & F\_J184239.7-040027  & 28.366  & 0.121   & 0.38  & 1.36  & 5.0  &  100  & No  \\ 
G028.37+00.07   (b) & F\_J184255.4-040150  & 28.376  & 0.053   & 0.51  & 3.34  & 5.0  &  310  & No  \\ 
G028.37+00.07   (e) & F\_J184300.2-040132  & 28.388  & 0.036   & 0.45  & 2.18  & 5.0  &  130  & No  \\ 
G028.37+00.07   (f) & F\_J184252.3-035956  & 28.403  & 0.064   & 0.43  & 3.72  & 5.0  &  4160 & No  \\ 
G028.53$-$00.25 (a) & F\_J184418.1-035938  & 28.563  & -0.232  & 0.41  & 12.8  & 5.7  &  4190 & Yes \\ 
G028.61$-$00.26 (a) & F\_J184428.1-035750  & 28.613  & -0.262  & 0.27  & 0.56  & 4.2  &  80   & Yes \\ 
G030.97$-$00.14 (a) & E\_J184821.9-014832  & 30.978  & -0.149  & 0.38  & 6.30  & 5.1  &  1270 & Yes \\ 
G031.03+00.26   (b) & F\_J184701.4-013438  & 31.023  & 0.261   & 0.29  & 2.97  & 6.6  &  520  & Yes \\ 
G031.03+00.26   (c) & F\_J184707.4-013432  & 31.036  & 0.239   & 0.29  & 1.32  & 6.6  &  160  & Yes \\ 
G031.03+00.26   (a) & F\_J184701.4-013314  & 31.043  & 0.273   & 0.31  & 3.43  & 6.6  &  460  & Yes \\ 
G031.38+00.29   (a) & F\_J184732.6-011338  & 31.393  & 0.299   & 0.38  & 6.96  & 6.6  &  2460 & No  \\ 
G031.97+00.07   (b) & F\_J184922.1-005038  & 31.943  & 0.074   & 0.42  & 0.88  & 6.9  &  420  & Yes \\ 
G031.97+00.07   (c) & F\_J184926.9-005002  & 31.961  & 0.061   & 0.35  & 0.82  & 6.9  &  490  & No  \\ 
G033.69$-$00.01 (e) & E\_J185248.6+003602  & 33.623  & -0.036  & 0.32  & 1.76  & 7.1  &  690  & No  \\ 
G033.69$-$00.01 (b) & E\_J185252.6+003832  & 33.663  & -0.032  & 0.38  & 1.60  & 7.1  &  620  & No  \\ 
G033.69$-$00.01 (c) & E\_J185253.8+004120  & 33.701  & -0.012  & 0.37  & 1.35  & 7.1  &  530  & No  \\ 
G033.69$-$00.01 (a) & E\_J185257.0+004302  & 33.743  & -0.012  & 0.38  & 7.05  & 7.1  &  2750 & Yes \\ 
G034.43+00.24   (a) & F\_J185318.9+012650  & 34.431  & 0.241   & 0.34  & 1.20  & 3.7  &  100  & Yes \\ 
G038.95$-$00.47 (a) & F\_J190407.5+050844  & 38.959  & -0.469  & 0.53  & 4.58  & 2.7  &  290  & Yes \\ 
G048.52$-$00.47 (a) & F\_J192207.4+133713  & 48.519  & -0.467  & 0.38  & 0.76  & 2.8  &  50   & No  \\ 
G048.65$-$00.29 (a) & F\_J192144.7+134925  & 48.658  & -0.289  & 0.34  & 1.97  & 2.5  &  100  & Yes \\ 
\hline\hline
\label{table:mass}
\end{tabular}
\end{minipage}
\end{table*}

\subsection{Detection rates}

\begin{figure}
\includegraphics[width=0.4\textwidth,angle=-90]{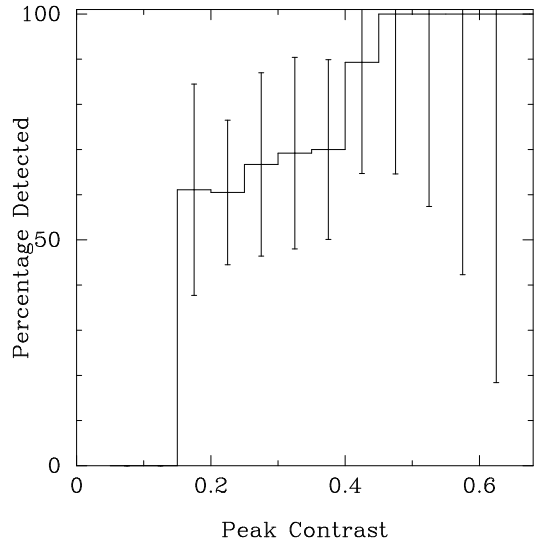}
\caption{Histogram of contrast value against detection rate (defined as IRDC cores with associated 850\,$\umu$m sources). Bin size used is 0.05.}
 \label{Hist:Reliability}
\end{figure}

We investigated the fraction of cores detected at 850\,$\umu$m versus peak contrast to identify any trends in the detection fraction of IRDC cores. We calculated the detection fraction by dividing the number of cores with 850\,$\umu$m detections by the total number of MSX IRDC cores within the SCUBA Legacy Catalogue at specific contrast values. Overall the fraction of IRDC cores detected by SCUBA is found to be 75\%. At low contrast values we detect over 60\% of the IRDC cores, whereas all IRDC cores are detected when the contrast is high as seen in Fig. \ref{Hist:Reliability}. The error bars in Fig. \ref{Hist:Reliability} were calculated by assuming an uncertainty of $\sqrt{N}$ where $N$ is the number of cores detected by SCUBA per bin and using propagation of errors.

\subsection{MIPSGAL 24\,$\umu$m sources associated with IRDCs detected at 850\,$\umu$m}
\label{subsection:SCUBAdetectedIRDCs}

\begin{figure}
 \includegraphics[width=0.4\textwidth,angle=-90]{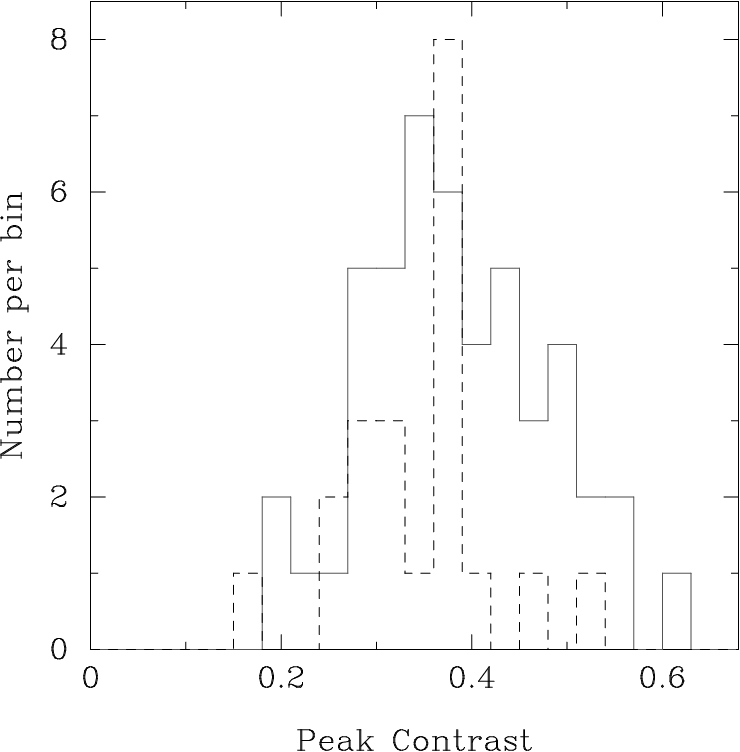}
\caption{Histogram showing the distribution of the 69 cores detected at 850\,$\umu$m within the MIPSGAL coverage area with (solid line) and without (dashed line) an embedded 24\,$\umu$m object at each Peak Contrast value. Bin size used is 0.03.}
 \label{Hist:embedded}
\end{figure}

We present a histogram of the 69 IRDC cores detected at 850\,$\umu$m within the MIPSGAL coverage area in Fig. \ref{Hist:embedded}. 48 of the IRDC cores were found to be positionally associated with one or more 24\,$\umu$m MIPSGAL sources and 21 were found not to be associated with any MIPSGAL sources

To investigate if the cores with and without embedded 24\,$\umu$m objects were drawn from the same population, a KS test was once again performed on their peak contrast distribution and it was found that they are highly likely to originate from the same population, with no distinct differences in their peak contrast distribution to a level of significance $>$\,99\%.

\section[]{Discussion}
\label{Section:Discussion}

\subsection{The reliability of the MSX IRDC catalogue}
\label{subsection:Reliability}

Originally, when the \cite{2006ApJ...639..227S} catalogue was published, an initial reliability of 82\% was reported for IRDCs with contrast values $>$\,0.25. This initial reliability was estimated for the large high contrast clouds by comparison to other source lists from MSX and ISO data (\citealt{2006ApJ...639..227S} and references there in). Later \cite{2008ApJ...680..349J} determined a reliability (against CS J=2--1 detections) for low contrast (0.2\,-\,0.4) objects of approximately 50\% increasing to almost 100\% at high contrasts ($>$\,0.6), with an overall reliability of $\sim$59\%. However the \cite{2008ApJ...680..349J} sample lacked very low contrast objects, the selection criteria used were peak contrasts $>$\,0.32 and angular sizes $>$\,42''. Expanding this estimate to the whole MSX IRDC catalogue \cite{2008ApJ...680..349J} stated that it was $>$\,50\% reliable for all contrasts. These estimates of reliability were obtained via molecular line spectroscopy of $^{13}$CO and CS data from \cite{2006ApJ...653.1325S} and \cite{2008ApJ...680..349J} respectively.

Our mean detection rate of IRDC cores with 850\,$\umu$m emission is 75\% (this value varies over a range of peak contrast values as seen in Fig. \ref{Hist:Reliability}) which is greater than the reliability of 50\% stated by \cite{2008ApJ...680..349J}. However this detection rate does not take into account the number of inconclusive matches that make up 19\% of the IRDC sample. The close correspondence of 850\,$\umu$m emission, CS 2--1 and $^{13}$CO detection rates for IRDCs places greater confidence in the high contrast \cite{2006ApJ...653.1325S} candidate IRDCs as true molecular clouds.

\subsection{Cores not detected at 850\,$\umu$m}
\label{Section:Discussion1}

Of the 205 cores within our sample 51 cores were not detected at 850\,$\umu$m. Those cores detected at 850\,$\umu$m were found to have higher peak contrast vales and column densities than those cores not detected at 850\,$\umu$m. The difference in peak contrast values can be seen in Fig. \ref{Hist:ColDen}. The median column density of the cores detected at 850\,$\umu$m is a factor of $\sim$1.6 times greater than the cores not detected at 850\,$\umu$m. A KS test of the peak contrast distribution could find no significant difference between the cores detected at 850\,$\umu$m and those that were not.

\begin{figure}
\includegraphics[width=0.4\textwidth,angle=-90]{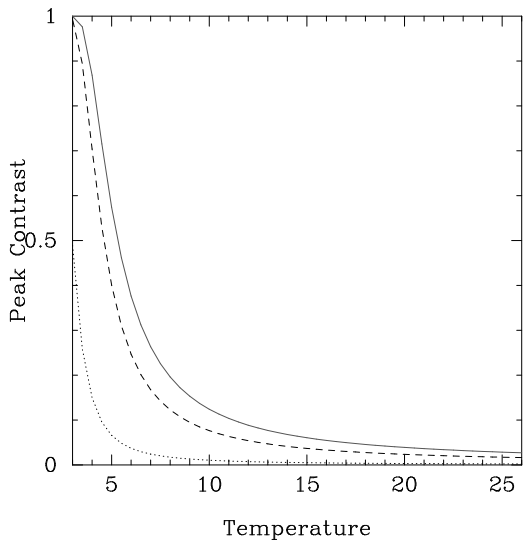}
\caption{A graph to show the sensitivity of SCUBA at 850\,$\umu$m (solid line) with respect to the detection of contrast values as a function of core temperature. The dashed line represents the predicted sensitivity of SASSy, which at 3\,$\sigma$ will equal 90\,mJy. The dotted line represents the predicted sensitivity of JPS, which at 3\,$\sigma$ will equal 12\,mJy.}
 \label{Graph:Sensitivity}
\end{figure}

Although the mid-infrared contrast value for a particular cloud should vary with the intensity of the background, it is possible to determine an estimate of the peak contrast sensitivity of SCUBA as a function of temperature by substituting equation \ref{equation:N8(H2)} into equation \ref{equation:N850(H2)}, using an 850\,$\umu$m flux limit of 3\,$\sigma$ (where $\sigma$ is the median rms sensitivity of the SCUBA Legacy Catalogue) for $F_{\nu}$ and rearranging for $C$:
\begin{equation}
C = 1 - exp \left( -\frac{F_{\nu} C_{\nu} \sigma_{\lambda}}{B_{\nu}(T) \pi (tan(B_{850}))^{2} 2 m_{H}} \right)
\label{equation:sensitivity}
\end{equation} The median rms sensitivity value at 850\,$\umu$m in the SCUBA Legacy Catalogue after applying our noise cut is 50\,mJy\,beam$^{-1}$. We plot the limiting contrast of the SCUBA Legacy Catalogue as a function of temperature, derived using equation \ref{equation:sensitivity}, in Fig. \ref{Graph:Sensitivity}. We also plot the forecast sensitivity of the SCUBA-2 Legacy Survey SASSy (the SCUBA-2 ``All Sky'' Survey; \citealt{2007arXiv0704.3202T}) and JPS (the JCMT Plane Survey; \citealt{2005prpl.conf.8370M}), which will be discussed further in Section \ref{section:SASSy}.

Fig. \ref{Graph:Sensitivity} shows that the median sensitivity of the SCUBA Legacy Catalogue would be sufficient to detect the majority if the \citet{2006ApJ...639..227S} IRDC cores if they have temperatures greater than 10\,K. At this temperature we would detect IRDC cores with peak contrast greater than 0.15. This corresponds to the approximate completeness limit of the \cite{2006ApJ...639..227S} catalogue, where the turnover in peak contrast occurs. In our sample of 205 IRDCs within the SCUBA Legacy Catalogue 94\% have peak contrast values greater than or equal to 0.15.

The IRDC cores that are not detected at 850\,$\umu$m are thus consistent with being low temperature, low column density cores but below the 0.1\,Jy\,pixel$^{-1}$ noise cut we applied to the SCUBA Legacy Catalogue. Fig. \ref{Graph:Sensitivity} shows that if they are true clouds they are likely to have temperatures less than 10\,K. Almost none were identified with 24\,$\umu$m MIPSGAL sources, corroborating our low temperature hypothesis and implying that they are either transient or potentially prestellar cores.

If we consider the cores not detected at 850\,$\umu$m, to be low temperature, low column density cores, this requires that they have a temperature lower than $\sim$10\,K, whereas the lowest contrast objects in the sample could have temperatures less than 14\,K. Typical temperatures for IRDCs range from 8--25\,K \citep{1998ApJ...508..721C,2002A&A...382..624T,2006A&A...450..569P}. However as we do not know how IRDC cores are arranged in temperature we cannot know if we ought to have detected the majority of these objects or not. We therefore cannot rule out the presence of a cold faint transient or prestellar population within the SCUBA non-detected sample, particularly at low contrast values where SCUBA is least sensitive. Recent theoretical models \citep{2007MNRAS.379.1390S} suggest that the temperature of prestellar cores may be lower than suspected ($\sim$5--10\,K), which would place the 850\,$\umu$m fluxes of the IRDC cores below our detection limit.

An alternative hypothesis is that a number of the cores not detected at 850\,$\umu$m are a result of the  absence of background mid-infrared emission rather than its extinction by intervening cold dust in an IRDC. In this case some the cores not detected at 850\,$\umu$m would be localised ``holes'' or local minima in the mid-infrared background masquerading as IRDCs. This possibility is more likely for the higher contrast cores not detected at 850\,$\umu$m, as our temperature constraints for these objects mean that they are less likely to be prestellar. Artefacts may also be present as a result of the background subtraction process, which would result in false IRDC detections particularly in regions of complex emission \citep{2006ApJ...653.1325S}. Without deeper submillimetre continuum or molecular line data it is difficult to satisfactorily determine whether an IRDC core identified by \citet{2006ApJ...639..227S} and not detected at 850\,$\umu$m is a true cloud, void or artefact.

\subsection{Cores detected at 850\,$\umu$m}

We identified 154 cores detected at 850\,$\umu$m within the area covered by the SCUBA Legacy Catalogue. These cores have higher peak contrast values (as seen in Fig. \ref{Hist:ColDen}) and column densities than those cores cores not detected at 850\,$\umu$m. Clearly as these objects are seen in submillimetre emission they are not voids or artefacts in the MSX contrast images. We determine estimates of the mass of the 33 cores in our sample with kinematic distances (see Section \ref{subsection:coldenmass}). The median mass of cores within the sample is 300\,M\,$_{\odot}$, with a minimum mass of 50\,M\,$_{\odot}$ and a maximum mass of 4,190\,M\,$_{\odot}$ (assuming a dust temperature of 15\,K). Our results are consistent with those of \citet{2006ApJ...641..389R}, who observed the 38 highest contrast clouds from \cite{2006ApJ...653.1325S} at 1.2\,mm, taking into account differences in sample selection, assumed temperature and in the measurement of integrated fluxes (\citealt{2006ApJ...641..389R} fit Gaussians to their sample whereas the SCUBA Legacy Catalogue uses Clumpfind).

Are these masses consistent with high mass star formation within the IRDCs? There is considerable uncertainty regarding the minimum mass core needed to form a high mass star, by considering the observed range in star formation efficiencies \citet{2006A&A...453.1003T} estimated that a core mass of at least 30--200\,M\,$_{\odot}$ would be required to form a 10\,M\,$_{\odot}$ star. Observed values for high mass star forming cores range from 720\,M\,$_{\odot}$ to 10$^{4}$\,M\,$_{\odot}$ \citep{2000A&A...357..637H,2002ApJS..143..469M}. Our sample of SCUBA detected IRDC cores falls at the lower end of this observed range of masses and is largely consistent with the estimate of the mass required for high mass star formation. We thus conclude that the masses of the IRDC cores in our sample are sufficient to support intermediate to high mass star formation.

\subsection{IRDC cores detected at 850\,$\umu$m without 24\,$\umu$m sources: could they be ``starless'' IRDCs?}
\label{section:starless}

Approximately two thirds of the IRDC cores detected at 850\,$\umu$m that are located within the MIPSGAL survey area are associated with embedded 24\,$\umu$m sources (48 cores or 69\% of the sample), as shown in Section \ref{subsection:SCUBAdetectedIRDCs}. We carried out KS tests of the peak contrast and column density distributions for SCUBA detected IRDC cores with and without associated 24\,$\umu$m sources. There is no evidence for the existence of two populations (see Fig. \ref{Hist:embedded}), which implies that the IRDC cores detected at 850\,$\umu$m with and without associated 24\,$\umu$m sources originate from the same column density population. We searched for any signs of correlation for the limited sample of those cores with known kinematic distances (see Table \ref{table:mass}) and did not find any correlation of the presence of an embedded object at 24\,$\umu$m with the mass of the IRDC core.

Given the similar properties of the cores with and without associated 24\,$\umu$m sources, the two types of core may be evolutionarily related. The cores that are without associated 24\,$\umu$m sources could represent an earlier ``starless'' evolutionary stage to the IRDC cores that have formed intermediate or high mass Young Stellar Objects and are associated with 24\,$\umu$m sources. A range of evolutionary stages have been observed in a handful of IRDCs \citep{2008arXiv0808.2973R}, which supports this hypothesis. In this picture the ``starless'' IRDC cores (i.e. those without associated 24\,$\umu$m sources) represent the cold quiescent initial conditions for high mass star formation as suggested by \citet{1998ApJ...508..721C} and \citet{2006ApJ...639..227S}. The SCUBA detected candidate IRDC cores with associated MIPSGAL 24\,$\umu$m sources would then represent a star forming population of IRDCs with embedded (proto) stellar objects, and so we refer to these as star forming IRDCs.

Two alternative explanations are that the starless IRDC cores detected at 850\,$\umu$m are sterile, possibly unbound, condensations that may never go on to form stars, or that they are forming stars, but with luminosities too low to be detected by MIPSGAL. In order to address the likelihood of the former explanation in detail, we would need to determine the virial masses and gravitational stability of a large sample of the starless IRDC cores (via additional spectroscopy). However we note that the Jeans Mass for a core of similar temperature and number density to the starless cores ($\sim$15\,K and 10$^{4}$\,cm$^{-3}$) is 20\,M$_{\odot}$. Decreasing the temperature or increasing the number density decrease the Jeans mass. The minimum mass of our sample of cores detected at 850\,$\umu$m (with or without associated 24\,$\umu$m sources) is 50\,M$_{\odot}$. Allowing for uncertainties in our derivation of the mass we thus conclude that it is unlikely that many of the IRDC cores fall below this Jeans Mass and so the majority of the IRDC cores detected at 850\,$\umu$m ought to at least have the potential for star formation.

This approach implicitly assumes that the IRDC cores are single gravitationally bound objects. If instead they are composed of numerous smaller cores fragmented below the scale of the JCMT beam then this conclusion may not apply. Higher resolution interferometry would be needed in this case (e.g. \citealt{2008arXiv0808.2973R}).

\begin{figure*}
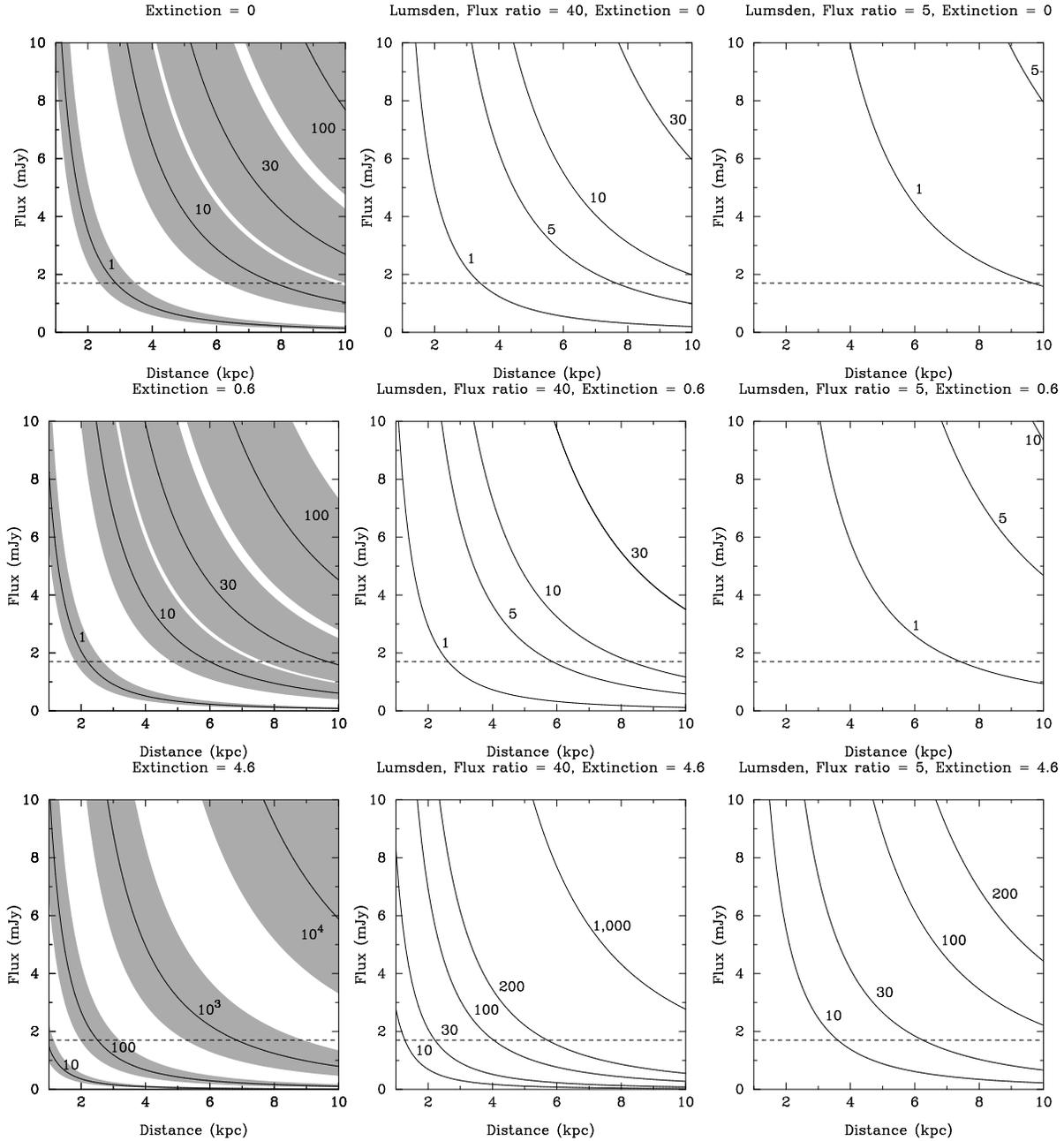

\includegraphics[width=0.32\textwidth,angle=-90]{dunhamE0range.eps}
\includegraphics[width=0.32\textwidth,angle=-90]{lumsdenR40E0.eps}
\includegraphics[width=0.32\textwidth,angle=-90]{lumsdenR5E0.eps}
\includegraphics[width=0.32\textwidth,angle=-90]{dunhamE06range.eps}
\includegraphics[width=0.32\textwidth,angle=-90]{lumsdenR40E06.eps}
\includegraphics[width=0.32\textwidth,angle=-90]{lumsdenR5E06.eps}
\includegraphics[width=0.32\textwidth,angle=-90]{dunhamE46range.eps}
\includegraphics[width=0.32\textwidth,angle=-90]{lumsdenR40E46.eps}
\includegraphics[width=0.32\textwidth,angle=-90]{lumsdenR5E46.eps}
\caption{Left Plots: based on relationship between flux and luminosity, from observations and theoretical models as produced in \protect\citet{dunham08}. The shaded regions depict the uncertainty in this relationship. Middle and Right Plots: based on flux luminosity relationship published by \protect\citet{lumsden02}. Lower and upper estimates of flux for particular luminosities are plotted in the middle and right plot respectively. This range in flux originates from an observed range in F$_{FIR}/$F$_{E}$ ratio which ranges from 5-40. The middle plot takes the value of 40 and the right plot takes the value of 5. Plots from top to bottom vary in A$_{24}$, from no extinction in the left to high extinction in the right plot. Labels in each plot denote the different luminosities, i.e. 1\,L$_{\odot}$, 10\,L$_{\odot}$, 30\,L$_{\odot}$ and 100\,L$_{\odot}$. The dashed horizontal line is the 5\,$\sigma$ point source sensitivity of MIPSGAL at 24\,$\umu$m.}
\label{Fig:dunhamlumsden}
\end{figure*}

To assess the likelihood of the latter explanation that the starless IRDC cores are forming stars with luminosities below the detection limit of MIPSGAL we need to determine the sensitivity of MIPSGAL to YSOs as a function of YSO luminosity. Fortunately a series of studies carried out by the Spitzer c2d\footnote{The Spitzer Space Telescope Legacy program “From Molecular Cores to Planet-Forming Disks” \citep{Evans2003}} survey team on nearby star forming regions allows us to characterise the 24\,$\umu$m flux to total internal YSO luminosity fairly well for low mass YSOs. \citet{dunham08} find an approximately linear relationship between the MIPS 24\,$\umu$m flux and total internal luminosity for low mass YSOs detected in the Spitzer c2d survey. This was found to be consistent with the predictions of radiative transfer models of low mass YSOs \citep{crapsi2008}. Unfortunately no such similar study exists for high mass YSOs and so we use the empirical mid to far-infrared flux relation of \citet{lumsden02}. We plot these relationships (corrected for an estimate of the average 24\,$\umu$m extinction of the IRDCs) for a range of YSO luminosities as a function of distance in Fig. \ref{Fig:dunhamlumsden}.

Uncertainties in Fig. \ref{Fig:dunhamlumsden} result from uncertainties in extinction and in the relationships between flux and luminosity as taken from \citet{lumsden02} and \citet{dunham08}. The main source of error in \citet{dunham08} is the uncertainty in the relationship between flux and luminosity which were obtained from observations as well as theoretical models. This uncertaintly is depicted in Fig. \ref{Fig:dunhamlumsden} by the shaded region in the plots on the left. The primary source of error in \citet{lumsden02} is due to the large range in observed flux ratio. This uncertainty is depicted in the middle and right plots with the lower estimate in the middle plots and the upper estimates in the plots on the right. From Fig. \ref{Fig:dunhamlumsden} we see that the relationships taken from \citet{lumsden02} and \citet{dunham08} are consistent with each other particularly the plots to the right and in the middle of Fig. \ref{Fig:dunhamlumsden}.

The quoted 5$\sigma$ point source sensitivity of MIPSGAL is 1.7\,mJy at 24$\umu$m \citep{MIPSdocumentation}. This limit will obviously vary from region to region depending upon the strength of the background emission and complexity of structures in the images. However the IRDCs presented here in general have low 24$\umu$m backgrounds and are relatively free from source crowding, hence we assume the given 5$\sigma$ limit is valid, which is show in Fig. \ref{Fig:dunhamlumsden} by a dashed horizontal line.

The effect of extinction from the environments surrounding the cores found without associated 24\,$\umu$m emission and from the material contained within the cores themselves lowers any observed radiation emitted from within. With these starless IRDC cores having high column densities ($\sim2\times10^{22}$\,cm$^{-2}$ from 8\,$\umu$m and 850\,$\umu$m data) and hence high opacities there are three possible situations we should consider for the flux that is observed: \emph{i)} there may be no extinction \emph{ii)} there may be a medium amount of extinction (taking this from the lower of the quoted column densities derived from the 8\,$\umu$m data) \emph{iii)} there may be high extinction (taking this from the higher of the quoted column densities as derived from the 850\,$\umu$m data). We derive values of visual extinction (A$_{v}$) for the two latter cases where extinction has an effect on the observed flux using the method given by \citet{RiekeLebofsky85}. The derived A$_{v}$ range between 11 and 94 which converts to a 24\,$\umu$m extinction (A$_{24}$) of 0.6 and 4.6 respectively. The effect of the differing extinction can be seen in Fig. \ref{Fig:dunhamlumsden} in which the top graphs are for no extinction, the middle graphs take the medium extinction case (A$_{24}=0.6$), and the bottom graphs assume high extinction (A$_{24}=4.6$). Errors on these estimates are on the order of $\sim$40\%, predominantly resulting from the uncertainty in the empirically derived conversion from N(H+H$_{2}$)/E(B--V) \citep{bohlin1978}.

At the typical distance of IRDCs within our Galaxy, which is 3.8\,kpc, we look at the sensitivity of MIPSGAL. In Fig. \ref{Fig:dunhamlumsden}, we see that in the `worst case scenario' MIPSGAL should be complete to embedded objects with luminosities above 100\,L$_{\odot}$ (when high mid-infrared extinction is considered; A$_{24}=4.6$, A$_{v}=94$). This completeness limit indicates the possibility of ruling out the presence of all but low mass YSOs. From \citet{iben1965} we find that in the protostellar stages of evolution a star with a final main sequence mass $<$\,2\,M$_{\odot}$ will never reach luminosities greater than 100\,L$_{\odot}$. Greater constraints on the column densities of these objects are required to allow us to have a better handle on the potential luminosities of these cores and determine if indeed they are a low mass population of cores.

\subsection{The lifetimes of starless and star forming IRDCs}
\label{section:SF}
From the previous section we have seen that the starless cores may be evolutionary related to those SCUBA detected cores associated with 24\,$\umu$m objects but with luminosities below the detection limit of MIPSGAL. Going one step further we may assume that the starless and star forming SCUBA detected IRDCs are at different evolutionary stages in the formation of high mass stars and so we can estimate the statistical lifetime of the starless quiescent phase. If each starless IRDC core evolves into a corresponding star forming IRDC core with one or more embedded 24\,$\umu$m source then the relative proportions of these objects in the sample should reflect the statistical life time of each type of object. In the sample of SCUBA detected IRDCs lying within the MIPSGAL survey area we find twice as many star forming IRDC cores with 24\,$\umu$m sources than starless IRDC cores. Thus if these two types of object do form an evolutionary sequence we would expect the starless phase to last half the lifetime of the star forming phase.

Estimates for the absolute lifetime of the embedded high mass star formation range from $10^{4}$--$10^{5}$ years for UCHII and embedded YSOs \citep{1989ApJ...340..265W,1989ApJS...69..831W}, a few 10$^{4}$ years for methanol masers \citep{2005MNRAS.360..153V}, and 1.2--7.9$\times10^{4}$ years for embedded high mass protostars \citep{2007A&A...476.1243M,2007A&A...463.1009P}. Taking the upper and lower bounds of these estimates we conclude that the starless phase of IRDCs, as an upper limit due to our assumption on evolution, last a few $10^{3}$--$10^{4}$ years. The proportions of starless and star forming IRDCs that we see are consistent with the proportion of massive infrared quiet high mass protostars to the massive protostellar stage as found in Cygnus X by \citet{2007A&A...476.1243M}. The lifetime of the starless IRDC phase is comparable to that found for the infrared quiet protostellar phase by \citet{2007A&A...476.1243M} who calculated the statistical lifetimes based on the proportion of massive infrared quiet high mass protostars to the massive protostellar stage as found in Cygnus X. Our statistical lifetime estimate for the starless IRDC phase is also approximately one to two orders of magnitude less than the extended lifetime of the low mass Class 0 evolutionary phase recently calculated by \citet{2008arXiv0811.1059E}.

Caution must be applied to comparing the estimated lifetime of starless IRDCs to the estimated lifetime of the high mass pre-stellar phase. As shown by \citet{2007A&A...476.1243M} for Cygnus X there are no high mass starless cores, which implies an age of less than 10$^{3}$ years for this phase. The ``starless'' IRDCs that we identify from their mid-infrared quietness may yet display other signs of star formation such as molecular outflows or methanol masers which would imply that they have a proto-stellar nature. This may be supported by the fact that their statistical lifetime is similar to the high mass proto-stellar phase identified by \citet{2007A&A...476.1243M}. Future investigations of these clouds to search for identifiers of high mass star formation are needed to estimate the lifetimes of the pre-stellar and proto-stellar phases found within these clouds. A number of forthcoming Galactic Plane surveys have these aims, such as the Methanol Multi Beam Survey (MMB; \citealt{2008arXiv0810.5201G}), the CORNISH\footnote{The Co-Ordinated Radio `N' Infrared Survey for High-mass star formation, \citet{2008ASPC..387..389P}} 5\,GHz survey \citep{2008ASPC..387..389P}, the Herschel Hi-Gal Survey \citep{2005prpl.conf.8163M}, and the JCMT Legacy Surveys SASSy and JPS \citep{2007arXiv0704.3202T,2005prpl.conf.8370M}, see Section \ref{section:SASSy}. However, regardless we have shown that the lifetime of a quiescent (before it shows evidence of activity in the mid-infrared) IRDC is approximately half of that spent in the embedded phase.

\subsection{Predictions and implications for Galactic Plane surveys}
\label{section:SASSy}

\begin{figure*}
\includegraphics[width=0.7\textwidth]{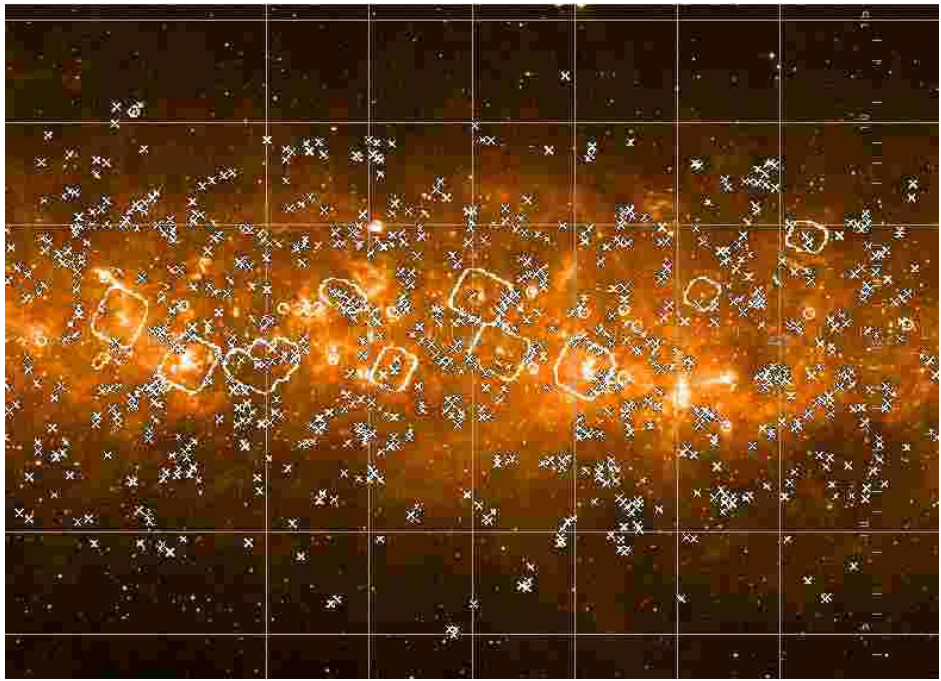}
\caption{Image of a region of the Galactic plane. MSX 8\,$\umu$m image with SCUBA coverage area contours overlaid. The crosses mark the locations of potential IRDC cores as catalogued by \protect\citet{2006ApJ...639..227S}.}
\label{pic:SASSy}
\end{figure*}

The astronomical comunity are planning a number of uniform and sensitive surveys of the Galactic Plane in the far infrared and sub millimetre that will detect a large number of the \citet{2006ApJ...639..227S} IRDC catalogue in emission. We use the results drawn form the SCUBA Legacy Catalogue to make predictions for the number of IRDCs that will be detected by four surveys in particular: SASSy, the SCUBA-2 ``All Sky'' Survey \citep{2007arXiv0704.3202T}, JPS, the JCMT Galactic Plane Survey \citep{2005prpl.conf.8370M}, Hi-GAL, the Herschel Infrared Galactic Plane Survey \citep{2005prpl.conf.8163M}, and ATLASGAL the APEX Telescope Large Area Survey of the Galaxy \citep{ATLASGAL}. Each of these surveys will cover much larger regions of the plane than the SCUBA Legacy Cataogue and will be both deeper and more uniform, resulting in a much more unbiased survey of IRDCs that is free from the targeted and non-uniform nature of the SCUBA Legacy Catalogue. Fig \ref{pic:SASSy} shows a region of the Galactic Plane with the coverage area of the SCUBA Legacy Catalogue and the positions of \cite{2006ApJ...639..227S} IRCDs, which clearly indicate the potential of these large area surveys to detect a large number of IRDCs.

As each of these surveys will detect IRDC cores by their emission rather than their extinction against the galactic mid-infrared background this means that they will also be sensitive to IRDCs located on the far side of the Galaxy that were not detected by \citet{2006ApJ...639..227S}. The forecast 1\,$\sigma$ sensitivities of HIGAL and JPS are 20\,mJy\,beam$^{-1}$ and 4\,mJy\,beam$^{-1}$ at 250\,$\umu$m and 850\,$\umu$m respectively, which are sufficient to detect cores of only a few tens of M$_{\odot}$ at 20\,kpc (assuming 20\,K dust with $\beta$=2 and a mass coefficient of 50\,g\,cm$^{-2}$). SASSy and ATLASGAL will have 1\,$\sigma$ sensitivities of 30\,mJy\,beam$^{-1}$ and 50-70\,mJy\,beam$^{-1}$ at 850\,$\umu$m which could detect cores of a few hundred M$_{\odot}$ out to 20\,kpc. Taking the masses of known IRDC cores into consideration each of these surveys has the potential to detect these objects at the far side of the Galaxy. In addition, as we have shown in Section \ref{Section:Discussion1}, the deeper surveys may find the low column density low temperature clouds that were not detected at 850\,$\umu$m in the SCUBA Legacy Catalogue. Thus as well as the increased number of detections resulting from surveying a larger area of the plane, we expect that the surveys will also detect a greater number of `IRDC cores' on the far side of the Galaxy and the colder population that we have not detected with SCUBA.

\begin{figure}
\includegraphics[width=0.4\textwidth,angle=-90]{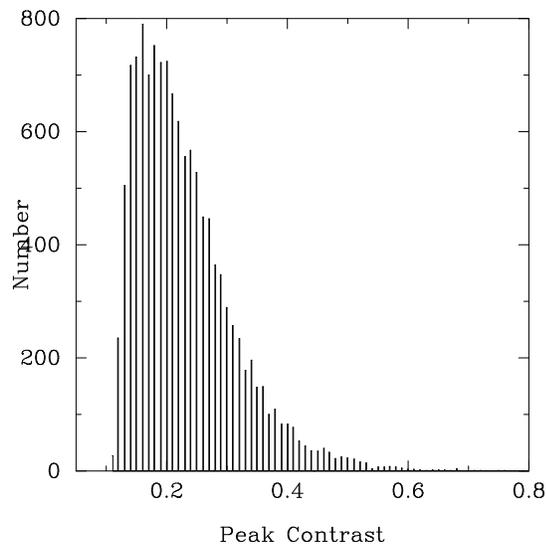}
\caption{Number distribution of IRDC cores with peak contrasts, values taken from \protect\citet{2006ApJ...639..227S}.}
 \label{Graph:CumulativeMSX}
\end{figure}

Estimating an upper limit to the number of IRDC cores that could be detected by the surveys is difficult. For the IRDC cores located on the far side of the Galaxy that have foreground emission preventing them being detected by \citet{2006ApJ...639..227S} we may estimate their number by geometric means and considering  the volume of the Galaxy probed by MSX. Following the argument presented by \citet{2006ApJ...641..389R} we estimate that the total number of IRDC cores in the Galaxy may be up to a factor of 3 greater than those detected by \citet{2006ApJ...639..227S}. To this number must be added an uncertain quantity of low column density cores whose intrinsic contrast falls below the \cite{2006ApJ...639..227S} criteria for detection but whose column density is great enough to be detected by the surveys (particularly Hi-GAL and JPS). We see from Fig. \ref{Graph:CumulativeMSX} that the steep turnover of IRDC cores at low contrast values may indicate that the catalogue is incomplete at low contrasts. Without further information on the general temperature distribution of IRDCs it is currently not possible to place firm limits on the number of such cores and so whilst we note that the deeper surveys will detect this colder population (and Hi-GAL will determine the temperature distribution of IRDC cores) we do not include them in our estimate.

Currently no information exists on the temperature distribution of IRDCs in general, as by the nature of their detection the estimated temperature for each cloud is an upper limit. This means that we cannot take the column densities estimated from the MSX 8\,$\umu$m data (as contained in Table \ref{table:classA}) and convert these into flux estimates, as the lack of temperature information renders these into rather loosely determined upper flux limits. In addition the large uncertainties in mass co-efficients, the 8\,$\umu$m extinction law and contamination from foreground emission introduce a considerable scatter between column densities derived from 8\,$\umu$m and 850\,$\umu$m (see Section \ref{subsection:coldenmass}). We thus estimate lower limits for the detection rate of IRDC cores within the surveys by using the SCUBA detection fraction shown in Section \ref{subsection:Reliability}. SASSY, JPS and Hi-GAL are deeper than the SCUBA Legacy Catalogue and so we expect these surveys to detect a greater fraction of IRDC cores, particularly at low contrast values where the surveys are more sensitive to low temperature low column density cores (see Fig. \ref{Graph:Sensitivity}). Without knowing the temperature distribution of IRDC cores it is impossible to determine exactly what this fraction is, but given the greater sensitivities of these surveys they ought to detect at least the fraction of IRDC cores that SCUBA did. The depth of ATLASGAL is similar to the 0.1\,Jy\,pixel$^{-1}$ noise cut that we applied to the SCUBA Legacy Catalogue and thus ATLASGAL should detect a similar fraction of IRDC cores from \citet{2006ApJ...639..227S}.

ATLASGAL will survey the inner third of the Galactic Plane ($|l|<60^{\circ}$ and $|b|<1.5^{\circ}$), within which there are 11,529 IRDC cores from the \citet{2006ApJ...639..227S} catalogue. Taking the SCUBA detection fraction of 75\% we predict that ATLASGAL will detect at least 8,600 IRDC cores. We scale this number by the geometric argument of \citet{2006ApJ...641..389R} to estimate the number of cores that ATLASGAL will detect on the far side of the Galaxy and hence  estimate that ATLASGAL may detect up to 26,000 IRDC cores. This is consistent with the preliminary results of the first 95 deg$^{2}$ of ATLASGAL which detects $\sim$6,000 sources, many of them infrared dark \citep{ATLASGAL}. The survey area of Hi-GAL again covers the inner third of the Galactic Plane but with a latitude range $|b|<1^{\circ}$. Of the 12,774 cores within the IRDC core catalogue by \citet{2006ApJ...639..227S} 10,644 IRDC cores are located within the Hi-GAL survey area. Scaling this to the detection fraction of SCUBA Hi-GAL will detect at least 8,000 IRDC cores. Again we use the geometric argument to estimate that within the entire Galaxy this number may increase to 24,000 cores. SASSy, covering  $0^{\circ}\leq l \leq245^{\circ}$ and $|b|\leq 5^{\circ}$ of the Galactic plane has 6,160 IRDC cores from the \cite{2006ApJ...639..227S} within the coverage area. Taking the 75\% detection fraction results in a lower estimate of 4,600 cores being obsevred. Again the number detected increases, when we consider geometric arguments, to 14,000. Finally JPS will survey two regions of the Galactic Plane, the GLIMPSE-N region ($10^{\circ}<l<65^{\circ}$ and $|b|\leq1^{\circ}$) and the FCRAO Outer Galaxy Survey region ($102.5^{\circ}<l<141.5^{\circ}$ and $|b|\leq1^{\circ}$). We see that 4,095 IRDC cores from \citet{2006ApJ...639..227S} are located within the coverage area. With a detection fraction of 75\% we expect a lower limit of 3,000 IRDC cores to be detected. With geometric arguments this number may increase to 9,000 IRDC cores. Although by number we see that the predicted numbers of IRDC cores that ATLASGAL and Hi-GAL are expected to return higher source counts than SASSy and JPS,  this is due to the larger area covered by these surveys. JPS and SASSy will however explore relatively unique parameter spaces. The high sensitivity JPS (1$\sigma$ $\sim$ 4\,mJy\,beam$^{-1}$ at 850 $\umu$m) will be ideal for identifying the most low temperature low column density IRDC cores. SASSy in contrast to the other surveys will have the benefit of observing greater latitudes of the Galactic Plane than any other survey and (as with JPS) will observe the outer Galaxy where low mid infrared backgrounds has restricted previous identifications of `IRDCs' due to their very nature.

\section[]{Summary and Conclusions}
\label{Section:Conclusions}

From positional cross matching of the IRDC catalogue produced by \citet{2006ApJ...639..227S} with the coverage area of the SCUBA Legacy Catalogue (as published by \citealt{2008ApJS..175..277D}) we have identified two populations of objects: candidate IRDC cores with and without associated 850\,$\umu$m emission. Column densities of these two populations were derived from the 8\,$\umu$m data by applying an extinction law to the peak contrast values (as defined by \citealt{2006ApJ...639..227S} based upon observations at 8\,$\umu$m). For those cores that were associated with 850\,$\umu$m emission column densities were also derived assuming a spherical geometry and the assumption of \citet{1983QJRAS..24..267H}. From our findings outlined within this paper we make the following conclusions:

\begin{enumerate}
\item 
We find 154 cores with 850\,$\umu$m detected emission and 51 cores without 850\,$\umu$m emission. Those cores associated with detectable 850\,$\umu$m emission had a median peak contrast value of 0.32, a median column density of 1.7$\times10^{22}$\,cm$^{-2}$ and a median mass of 300\,M$_{\odot}$. We found that the overall detection fraction of IRDC cores with 850\,$\umu$m emission is 75\%, as a lower limit which is in good agreement with the CS detection of \citet{2008ApJ...680..349J}.
\item
Those cores without 850\,$\umu$m emission are found to have no significant difference in peak contrast distribution than those cores detected at 850\,$\umu$m. These cores are likely to be population of low temperature low column density transient or prestellar cores. However, a small number of these cores could also be ``holes'' in the background mid-infrared continuum emission or artefacts as a result of the identification procedure. Further observations of the cores not detected at 850\,$\umu$m, either deeper sub millimetre continuum data or molecular line data, are required to yield insight into the true nature of these objects.
\item
On the nature of those cores detected at 850\,$\umu$m, we find that their range in masses ($50-4,190$\,M$_{\odot}$) are consistent with the lower mass end range observed in high mass star forming regions. 69\% of those cores detected at 850\,$\umu$m lying within the MIPSGAL survey area are associated with an embedded object at 24\,$\umu$m. A KS test gave no indication for the existence of two populations. This could suggest these cores are related evolutionarily. Those cores detected at 850\,$\umu$m without 24\,$\umu$m sources could be ``starless'' IRDCs or they may be forming stars but with luminosities too low to be detected. An alternative explanation for their origins are that they are unbound condensations that may never go on to form stars. To make more detailed conclusions about the nature of the SCUBA detected cores and their embedded mid-IR sources requires a deeper understanding of their physical properties from follow up molecular line mapping.
\item
Based on the assumption that the ``starless'' and star forming cores are related evolutionarily we derive an upper limit of $10^{3}-10^{4}$\,years for the lifetimes of starless IRDC cores. This lifetime is found to be comparable to the infrared quiet protostellar phase by \citet{2007A&A...476.1243M} and is approximately one to two orders of magnitude less than the extended lifetime of the low mass Class 0 evolutionary phase recently calculated by \citet{2008arXiv0811.1059E}.
\item
Based on SCUBA detection rates found, we make a conservative prediction to a lower limit of the number of IRDC cores that the Galactic Plane surveys ATLASGAL, Hi-GAL, SASSy and JPS will potentially detect : 8,600, 8,000, 4,600 and 3,000 cores respectively. If we apply geometric arguments to these values to scale to the number of such cores in the far Galaxy \citep{2006ApJ...641..389R} we see that ATLASGAL, Hi-GAL, SASSy and JPS have the potential to observe up to 26,000, 24,000, 14,000 and 9,000 infrared dark cores respectively throughout the Galaxy.
\end{enumerate}

We are now entering into an exciting time for sub millimetre and far infrared astronomy with the advent of Herschel and SCUBA-2. These two instruments will push the observational investigations of IRDCs, and in turn they will yield fresh insight into the role they may play in massive star formation 

\section*{Acknowledgments}

We thank the STFC and the University of Hertfordshire for support and an anonymous referee for a number of useful comments that substantially improved this work. We also thank James Fi Francesco for providing a file of all SCUBA Legacy Catalogue noise values. This research has made use of NASA's Astrophysics Data System. We would also like to acknowledge the JCMT (The James Clerk Maxwell Telescope is operated by The Joint Astronomy Centre on behalf of the Science and Technology Facilities Council of the United Kingdom, the Netherlands Organisation for Scientific Research, and the National Research Council of Canada), MIPSGAL and Midcourse Space Experiment. This research used the facilities of the Canadian Astronomy Data Centre operated by the National Research Council of Canada with the support of the Canadian Space Agency.

\bibliographystyle{mn2e}
\bibliography{bibliography}

\bsp

\label{lastpage}
\end{document}